\documentclass[journal]{IEEEtran}
\usepackage{rotating}
\usepackage{amsmath}
\usepackage{graphicx}
\usepackage{subfigure}
\usepackage{algpseudocode}
\usepackage{algorithm}
\usepackage{algorithmicx}
\usepackage{algpseudocode}
\usepackage{amssymb}
\usepackage{float}
 \usepackage{verbatim}
\usepackage{multirow}
\usepackage{makecell}
\usepackage{hyperref}
\usepackage{graphics} 
\usepackage{diagbox}
\usepackage{color}
\usepackage{multirow,booktabs}
\usepackage{adjustbox}

\usepackage[table]{xcolor} 
\usepackage{cleveref}

\usepackage[numbers,sort&compress]{natbib}

\begin{document}

\title{Deep Learning-Based Wideband Spectrum Sensing with Dual-Representation Inputs and Subband Shuffling Augmentation}

\author{Shilian~Zheng,
        Zhihao~Ye,
        Luxin~Zhang,
        Keqiang~Yue,
        and Zhijin~Zhao
    


\thanks{S. Zheng is with the College of Communication Engineering, Hangzhou Dianzi University, Hangzhou 310018, China. He is also with National Key Laboratory of Electromagnetic Space Security,
Jiaxing 314033, China (e-mail: lianshizheng@126.com).}

\thanks{Z. Ye, and Z. Zhao are with the College of Communication Engineering, Hangzhou Dianzi University, Hangzhou 310018, China (e-mail: 232080132@hdu.edu.cn; zhaozj03@hdu.edu.cn).}

\thanks{K. Yue is with the Key Laboratory of RF Circuits and Systems, Ministry of Education, Hangzhou Dianzi University, Hangzhou, Zhejiang 310005, China (e-mail: kqyue@hdu.edu.cn).}


\thanks{L. Zhang is with {{National Key Laboratory of Electromagnetic Space Security}}, Jiaxing 314033, China (e-mail: lxzhangMr@126.com).}

}

\maketitle

\begin{abstract}
The widespread adoption of mobile communication technology has led to a severe shortage of spectrum resources, driving the development of cognitive radio technologies aimed at improving spectrum utilization, with spectrum sensing being the key enabler. This paper presents a novel deep learning-based wideband spectrum sensing framework that leverages multi-taper power spectral inputs to achieve high-precision and sample-efficient sensing. To enhance sensing accuracy, we incorporate a feature fusion strategy that combines multiple power spectrum representations. To tackle the challenge of limited sample sizes, we propose two data augmentation techniques designed to expand the training set and improve the network's detection probability. Comprehensive simulation results demonstrate that our method outperforms existing approaches, particularly in low signal-to-noise ratio conditions, achieving higher detection probabilities and lower false alarm rates. The method also exhibits strong robustness across various scenarios, highlighting its significant potential for practical applications in wireless communication systems.
\end{abstract}


\section{Introduction}
\IEEEPARstart{W}{ith} the rapid development of mobile communication technology, the scarcity of spectrum resources has become a critical factor restricting the development of wireless communication systems \cite{1,2,3,4}. The commercialization of 5G technology and the research on 6G technology indicate that future wireless communication will face a broader range of application scenarios and higher data transmission demands. However, the traditional static spectrum allocation method can no longer meet the increasing spectrum demand, which has driven the development of Dynamic Spectrum Access (DSA) technology. Cognitive Radio (CR), as the core of DSA, aims to achieve efficient spectrum utilization through intelligent spectrum sensing, decision-making, and access strategies. Spectrum sensing, as the first step of CR technology, directly affects spectrum utilization and communication quality \cite{5,6}. Therefore, the research on wideband spectrum sensing technology has important theoretical and practical significance.

Traditional wideband spectrum sensing methods are generally divided into two categories: those based on Nyquist sampling and those based on sub-Nyquist sampling. 
Wideband spectrum sensing techniques based on Nyquist sampling often divide the wideband spectrum into multiple subbands, performing spectrum sensing by sequentially detecting each subband \cite{7,8,9,10,11}. For example, Mustapha et al. \cite{7} proposed using narrowband filters at the radio frequency front end to scan each subband sequentially. Although this approach is simple and easy to implement, it suffers from prolonged detection cycles. To address this limitation, some researchers employed multiple parallel frequency-shifting band-pass filter banks to divide the wideband signal into several narrowband signals \cite{8,9}. While this method reduces detection delay, it introduces significant complexity due to the intricate parallel filter structures and the requirement for numerous RF components, increasing both cost and implementation difficulty. Another approach involves estimating the Power Spectral Density (PSD) of the signal to extract the frequency positions of subbands and using energy detection to identify the presence of primary user signals \cite{10,11}. While this reduces computational complexity to some extent, it struggles with the noise power uncertainty problem. To address local irregularities in the signal’s frequency domain edges, wavelet transform-based wideband spectrum detection methods have been proposed \cite{12,13}. These methods mitigate the noise power uncertainty issue to some degree but remain challenged by low Signal-to-Noise Ratio (SNR) scenarios.


Sub-Nyquist sampling-based wideband spectrum sensing exploits the sparsity of wideband signals in the frequency domain, and achieves this through sampling at rates lower than the Nyquist rate. The introduction of Compressive Sensing (CS) theory \cite{14,15,16} has provided an effective solution for this technique. CS leverages frequency domain sparsity, making it possible to obtain wideband spectrum information at lower sampling rates. Current sub-Nyquist sampling mechanisms fall into three categories: Analog-to-Information Converter (AIC) \cite{17}, Modulated Wideband Converter (MWC) \cite{18}, and Multi-Coset Sampling (MCS) \cite{19}. In spectrum sensing methods based on compressed sensing, Qin et al. effectively recovered the original wideband signal from undersampled data for spectrum detection \cite{20,21}. Other approaches directly recovered the signal’s PSD for spectrum sensing, bypassing full signal reconstruction and simplifying the process \cite{22,23,24,25}. Although these algorithms reduce sampling requirements, they exhibit high computational complexity and rely heavily on the quality of signal reconstruction for optimal performance.

To enhance the performance of wideband spectrum sensing, researchers have increasingly turned to deep learning-based methods. Building on its remarkable success in fields like computer vision, deep learning has also demonstrated its effectiveness in addressing challenges in wireless signal processing \cite{26,27,28}. Numerous studies have investigated deep learning for spectrum sensing, covering both narrowband \cite{29,30} and wideband spectrum sensing \cite{31,32,33,34}. These deep learning-based methods can be broadly categorized into two types: target detection-based wideband spectrum sensing and multi-label classification-based wideband spectrum sensing. Target detection-based wideband spectrum sensing often utilizes image information for detection. For example, Pan et al. \cite{31} reframed the problem as an image processing task by employing algorithms to derive the cyclic spectrum of Orthogonal Frequency Division Multiplexing (OFDM) signals. By analyzing cyclic autocorrelation properties and converting them into grayscale images, a spectrum sensing framework based on the LeNet-5 model was constructed. Similarly, Gerstacker et al. \cite{32} used YOLO to detect processed time-frequency images and identify spectrum holes. In \cite{34}, researchers expanded on this approach by incorporating multi-feature inputs such as time-frequency images, signal vector graphs, and eye diagrams into a neural network, improving detection accuracy. Multi-classification model-based wideband spectrum sensing methods transform wideband spectrum sensing into a classification problem \cite{35,36,37,38}. Yan et al. \cite{35}, for instance, utilized energy measurements from multiple Secondary Users (SUs) as classification features input to a Convolutional Neural Network (CNN), effectively reducing online classification latency. Tian et al. \cite{37} developed the DeepSense network, which processes raw In-phase and Quadrature (IQ) samples directly to identify multiple spectrum holes simultaneously. This approach significantly improves detection performance and reduces latency compared to traditional energy detection methods. Building on DeepSense, Mei et al. \cite{38} introduced the ParallelCNN model, which splits raw IQ samples into two streams processed by parallel networks, further reducing the complexity of the network.

However, target detection-based wideband spectrum sensing relies on high-quality time-frequency image representations, which are challenging to generate and exhibit significantly degraded performance under low SNR conditions. On the other hand, multi-classification model-based wideband spectrum sensing methods that use raw IQ samples as input simplify implementation but fail to fully leverage the rich spectral information crucial for spectrum sensing. The lack of explicit spectral inputs limits the models' ability to learn and capture critical frequency-domain features, significantly impacting detection accuracy, particularly under low SNR conditions. Furthermore, most existing deep learning-based approaches require extensive training data to effectively optimize CNN parameters and mitigate overfitting risks. Their limited performance in few-shot scenarios remains a significant challenge, highlighting the need for further research to address this gap and enhance their applicability in data-scarce environments.


To address the challenges mentioned above, in this paper we propose a novel approach based on multiple power spectral inputs. We introduce an efficient deep learning framework designed for high-precision and sample-efficient wideband spectrum sensing. Our method preprocesses signals to extract multiple power spectral representations, which serve as inputs for a CNN. Additionally, we develop an efficient feature fusion network and implement data augmentation techniques to ensure high accuracy, even in scenarios with limited sample sizes. Compared to existing methods such as ParallelCNN and DeepSense, the proposed framework significantly enhances the performance of wideband spectrum sensing, effectively overcoming previous limitations. The main contributions of this paper are as follows:
\begin{itemize}

\item We leverage the PSD obtained through the Multi-Taper Method (MTM) as input to a CNN for wideband spectrum sensing. Compared to IQ-based inputs, the PSD is more sensitive to spectral variations, which enhances sensing accuracy without increasing network complexity, while also effectively controlling the false alarm probability. 

\item We propose a wideband spectrum sensing framework based on dual-representation power spectrum inputs. This framework uses two power spectrum representations of the signal, obtained through different preprocessing techniques, as inputs to a CNN classifier. Compared to traditional single-representation methods, the dual-representation input significantly enhances the network's feature extraction ability, improving spectrum sensing accuracy.
\item We propose two data augmentation methods tailored for wideband spectrum sensing in few-shot scenarios. These techniques involve shuffling the spectrum either within subbands or across subbands, generating additional training samples that improving the network’s detection performance in data-limited environments.

\item We evaluate the proposed method's performance in terms of detection probability and false alarm rate in various scenarios. Results demonstrate that our approach outperforms traditional IQ-based and Periodogram (PG)-based algorithms, offering higher detection probability and lower false alarm rates, especially under low SNR conditions. The framework also shows strong generalization and resilience across various channel environments, making it a promising solution for efficient spectrum utilization in cognitive radio networks.


\end{itemize}

The rest of this paper is organized as follows. In Sec. \ref{sec1}, we present the problem model for wideband spectrum sensing. In Sec. \ref{sec2}, we provide a detailed exposition of the DSFF method proposed in this paper. In Sec. \ref{sec3}, we compare the spectrum sensing performance of the DSFF method with existing approaches.  Finally, we summarize the content of this paper in Sec. \ref{sec4}.

\section{Problem Formulation}
\label{sec1}
\begin{figure}[t]
    \centering 
    \includegraphics[width=8cm]{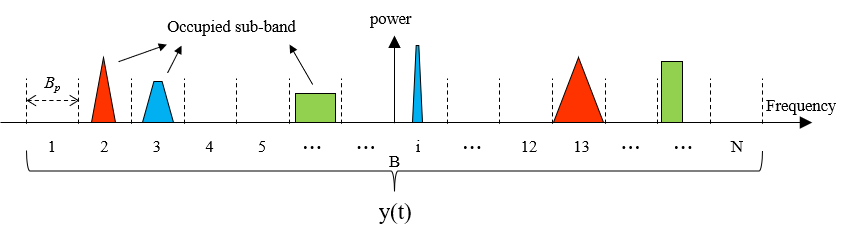}  
    \caption{Wideband signal scenario.}
    \label{fig1}
\end{figure}

In the allocation of primary spectrum resources, the entire bandwidth is divided into multiple sub-bands, each of which is allocated to different primary users. Assuming the system's total bandwidth is $B$, it is divided into $N$ sub-bands, as illustrated in Fig. \ref{fig1}. In the context of spectrum sensing, the signal received by secondary users is typically a superposition of signals from multiple sub-bands, represented as:
\begin{equation}
y(t)=\sum_{i=1}^Ns_i(t)+w(t),
\label{eq1}
\end{equation}where $s_{i}(t)$ represents the transmission signal of the primary user on each sub-band, and if the $i$-th sub-band is not occupied by a primary user, then $s_{i}(t)=0$; $w(t)$ represents the Additive White Gaussian Noise (AWGN). Unlike narrowband spectrum sensing, the goal of wideband spectrum sensing is to simultaneously monitor and analyze each sub-band within a wide frequency range to identify and utilize unoccupied spectrum resources.

We model wideband spectrum sensing as a combination of multiple binary hypothesis testing problems, each of which can be expressed as:
\begin{equation}
\begin{aligned}&H_{0}:Y_{i}(f)=W(f),
\\&H_{1}:Y_{i}(f)=S_{i}(f)+W(f),\end{aligned}
\label{eq2}
\end{equation}
where $Y_{i}(f)$ is the spectrum of the received signal on the $i$-th sub-band, $S_{i}(f)$ is the spectrum of the primary user's transmitted signal, $W(f)$ is the spectrum of AWGN, $H_0$ represents the absence of the primary user signal, and $H_1$ represents the presence of the primary user signal. The secondary user processes the received signal to obtain a decision statistic $\gamma_i$, which is then compared to a predetermined threshold $\lambda_i$ to determine the usage state of the target subband:
\begin{equation}
\gamma_i \underset{H_0}{\overset{H_1}{\gtrless}} \lambda_i.
\label{eq3}
\end{equation}


To comprehensively evaluate the overall performance of wideband spectrum sensing, we need to integrate the detection probabilities and false alarm probabilities of all sub-bands. In the case of imbalanced data, the micro-averaging method is more suitable for wideband spectrum sensing. Micro-averaging accumulates the detection results of each sub-band to calculate the overall detection probability and false alarm probability, thereby providing a more accurate reflection of the detector's overall performance on imbalanced datasets. The formulas for calculating the micro-averaged detection probability and false alarm probability are as follows:

\begin{equation}
P_d=\frac{\sum_{i=1}^N(\mathrm{TP})_i}{\sum_{i=1}^N(\mathrm{TP})_i+(\mathrm{FN})_i},
\label{eq5}
\end{equation}
\begin{equation}
P_{f}=\frac{\sum_{i=1}^N(\mathrm{FP})_i}{\sum_{i=1}^N(\mathrm{FP})_i+(\mathrm{TN})_i},
   \label{eq6}
\end{equation}
where $P_d$ is the micro-averaged detection probability, $P_f$ is the micro-averaged false alarm probability, and $TP_i=(H_1|H_1)_i$, $FP_i=(H_1|H_0)_i$, $TN_i=(H_0|H_0)_i$, $FN_i=(H_0|H_1)_i$ represent the number of true positives, false positives, true negatives, and false negatives, respectively, in the $i$-th sub-band. 


\section{Methodology}
\label{sec2}

\subsection{Overall Framework}

\begin{figure*}[tp]
    \centering
    \includegraphics[width=0.8 \textwidth]{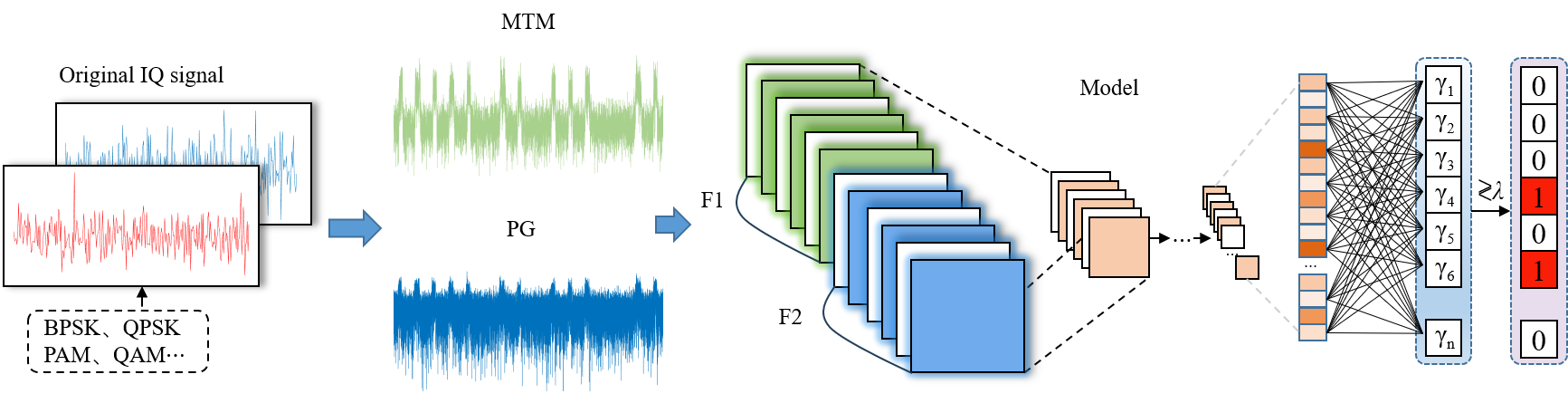}
    \caption{Wideband spectrum sensing structure.}
    \label{fig.structure}
\end{figure*}

We propose a deep learning-based wideband spectrum sensing framework, which incorporates power spectrum representation fusion, as illustrated in Fig. \ref{fig.structure}. SUs first compute two types of power spectrum representations from the received IQ signal: one based on MTM and another based on PG. These two representations are then processed by separate feature extraction modules, F1 and F2, which employ convolutional layers to learn deep feature maps tailored to each input. The extracted feature maps are subsequently fused through a fusion module to combine complementary information from both representations effectively. In the final stage, the fused features are passed to multiple binary classifiers, with each classifier dedicated to one subband. These classifiers generate decision metrics for the usage status of their corresponding subbands. The decision metrics are then compared with predefined thresholds to determine whether each subband is occupied by a primary user or not. The framework leverages the complementary strengths of both representations, enabling robust and accurate spectrum sensing under varying signal conditions.



\subsection{Power Spectrum Input}
The method designed in this paper combines two types of power spectrum representations computed using the PG and the MTM. Assume $y[m]$ represents a discrete-time sequence, where $m=0,1,2,\dots,M-1$ and $M$ is the length of the signal. The calculation of the power spectrum using the PG is as follows: First, the Discrete Fourier Transform (DFT) of the signal is computed:
\begin{equation}
Y[k]=\sum_{m=0}^{M-1}y[m]\cdot e^{-j\frac{2\pi}{M} km}
\label{eqPG},
\end{equation}
where $Y[k]$ is the $k$-th DFT coefficient of the sequence $y[m]$. Next, by shifting the spectrum center, the zero frequency point is placed at the center of the spectrum. The power spectrum $ \hat{P}[k]$ is the squared magnitude of the DFT coefficient: 
\begin{equation}
     \hat{P}[k]=|Y[k]|^2
    \label{eq7}.
\end{equation}

The calculation of the power spectrum using the MTM can be represented as:
\begin{equation}
Y_l[k]=\sum_{m=0}^{M-1}y[m]v_m^{(l)}\cdot e^{-j\frac{2\pi}{M} km},
\label{eq8}
\end{equation}
where $v_{m}^{(i)}$ represents the Slepian orthogonal sequence, $l=0,1,2,\dots,L-1$, and $L$ represents the degrees of freedom that control the variance of the multi-taper spectral estimate. The time-bandwidth product $C_o=MB$ determines the maximum number of orthogonal tapers, subject to the following constraint:
\begin{equation}
   L\leq\left \lfloor 2MB \right \rfloor ,
   \label{eq9}
\end{equation}where $\left \lfloor \cdot \right \rfloor$ denotes the floor function, which rounds a real number down to the nearest integer. The Slepian sequences have specific energy concentration properties, and their distribution in time and frequency depends on the choice of the window. For each Slepian sequence $v_m^{(l)}$, its energy $E_l$ can be expressed as: 
\begin{equation}
E_{l}=\sum_{m=0}^{M-1}\left|v_{m}^{(l)}\right|^{2}.
\end{equation}

The power spectrum for each taper $  \hat{P}_l[k]$ is computed as the squared magnitude of the frequency-domain representation:
\begin{equation}
     \hat{P}_l[k]=|Y_l[k]|^2
    \label{eq11}.
\end{equation}
The energy of the power spectrum $ \hat{P}_l[k]$ is concentrated within a resolution bandwidth of $2B$. To obtain the final estimate of the power spectrum, we averaged the values of all tapers. The weighted average of the spectral estimates is given by:
\begin{equation}
    \hat{P}[k]=\frac{\sum_{l=0}^{L-1} w_{l}\left|Y_{l}[k]\right|^{2}}{\sum_{l=0}^{L-1} w_{l}},
    \label{eq12}
\end{equation}
the weights $w_l$ are typically based on the energy $E_l$ of each Slepian sequence. A common approach is to normalize the energies of the tapers so that the weights sum to one, ensuring that the contribution from each taper is proportional to its energy. The weights are calculated as: 
\begin{equation}
    w_l=\frac{E_l}{\sum_{l=0}^{L-1}E_l} .
\end{equation}


MTM significantly reduces the variance of spectral estimation by utilizing multiple tapers, while the PG tends to be more susceptible to noise interference due to its higher variance. Additionally, the Slepian sequences used in MTM maximize energy concentration within a specified bandwidth, enabling more accurate capture of the signal's spectral characteristics. As a result, MTM exhibits stronger robustness to noise and interference, making it especially suitable for signal processing in low SNR environments. The fusion of MTM and PG effectively combines the complementary strengths of both methods. MTM provides low-variance, noise-resistant spectral estimation, while PG preserves the original spectral structure of the signal more faithfully. By integrating these two approaches, the proposed framework can simultaneously capture detailed spectral features (via PG) and enhanced energy concentration (via MTM).




 \subsection{Multi-Label Classification}

Multi-Label Classification (MLC) addresses the task of assigning multiple labels to a single sample. Unlike traditional single-label classification, MLC allows each sample to be associated with a set of classes, potentially containing multiple labels. 
In MLC tasks, there are various strategies to handle the multi-label attributes of samples. The First-Order Strategy (FOS) decomposes the MLC problem into multiple independent binary classification tasks by converting the original label $Z$ into a sequence of 0 and 1. Each task determines whether a specific label $z_i$ exists within the sample. 

The wideband spectrum sensing model can be viewed as a combination of multiple binary hypothesis models, similar to the FOS in MLC. Therefore, in our method we use an MLC model to handle the wideband spectrum sensing problem. In this framework, the number of classes in the MLC task corresponds to the number of subbands in wideband spectrum sensing. Each subband is analogous to a category label in MLC, and binary cross-entropy is used as the loss function for the CNN. The expression of the loss function is:

\begin{equation}
    \begin{split}
        Loss_{BCE}(\gamma,Z) &= -\frac{1}{C*N}\sum_{c=1}^{C}\sum_{i=1}^{N} \left[ Z_i^c \log(\gamma_i^c) \right. \\
        &\hfill + \left. (1 - Z_i^c) \log(1 - \gamma_i^c) \right],
    \end{split}
    \label{eqloss}
\end{equation}


where $C$ represents the total number of samples, and $\gamma_i^c$ and $Z_i^c$ denote the decision statistic and true label, respectively, for the $i$-th subband of the $c$-th sample. The overall loss is averaged across all subbands to evaluate the CNN's learning effectiveness for each subband's information.

During neural network training, we use the Adam optimizer based on Stochastic Gradient Descent (SGD) to iteratively update model parameters and minimize the loss function. To prevent overfitting, early stopping is applied: if accuracy does not improve over a certain period, the learning rate is reduced to 0.1 times its original value. If accuracy still does not improve over a longer period, training is stopped. These strategies enhance the model’s generalization and training effectiveness. 

\subsection{Network Structure}

\begin{figure}[t]
    \centering 
    \includegraphics[width=8cm]{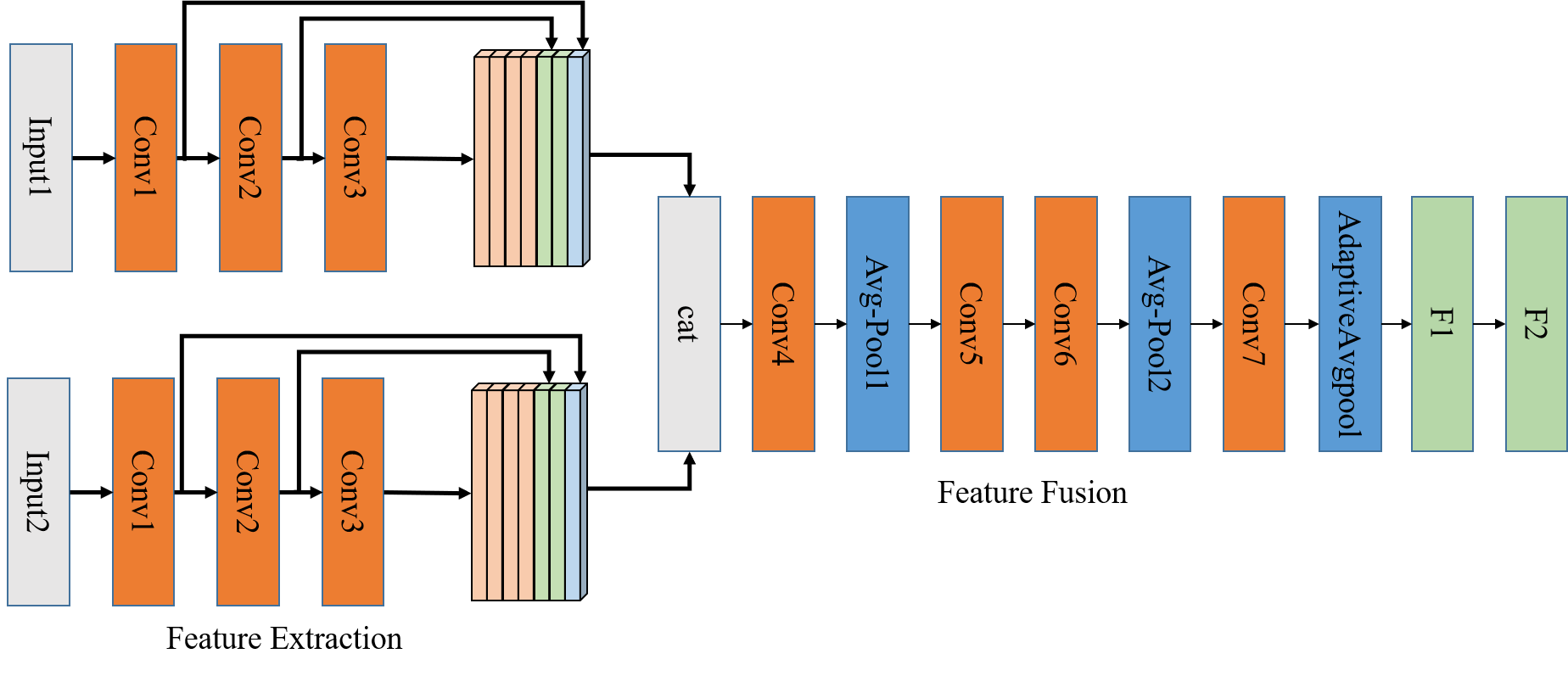}  
    \caption{DSFF Network Structure Diagram.}
    \label{fig3}
\end{figure}

In our proposed framework, we design a feature extraction and fusion network named Dual-Stream Feature Fusion Network (DSFF). The network includes a feature extraction module and a feature fusion module, aiming to effectively extract information from the two inputs. Fig. \ref{fig3} shows the structure diagram of the DSFF network, and TABLE \ref{CNN_data} summarizes the hyperparameters of each module in the network in detail. 

\begin{table}[t]
    \centering
        \caption{Summary of CNN dimensions}
    \begin{tabular}{ccccc}
    \hline
         Layer(Activation)&  Filter&  Stride&  padding& Length\\
         \hline
         Input&  &  &  & 32768
\\
         Conv1&  4*7&  1&  3& 32768
\\
         Conv2&  4*5&  1&  2& 32768
\\
         Conv3&  8*3&  1&  1& 32768
\\
         Conv4&  32*5&  2&  & 16382
\\
         Avg-Pool1&  2&  2&  & 8191
\\
         Conv5&  64*1&  1&  & 8191
\\
         
         Conv6&  64*5&  2&  & 4094
\\
         Avg-Pool2&  2&  2&  & 2047
\\
         Conv7&  32*1&  1&  & 2047
\\
         AdaptiveAvgpool+Flatten&  256&  &  & 8192
\\
         L1&  64(8192)&  &  & 64
\\
         L2&  16(64)&  &  & 16
\\
\hline
    \end{tabular}
    \label{CNN_data}
\end{table}

In CNN, the receptive field of the convolution operation is limited, which often makes it challenging for the network to capture the overall features of the input signal on a single scale. Therefore, in the feature extraction module, we apply convolution with kernels of different scales to capture feature information from various receptive fields. Through this multi-scale convolution design, the features extracted by each convolution kernel are combined into a comprehensive feature representation, which serves as the input to the feature fusion module to obtain a more complete feature expression.

To improve the network's convergence speed and enhance training stability, each convolutional layer is followed by ``BatchNorm" and ``LeakyReLU" operations. Specifically, ``BatchNorm" refers to batch normalization, which ensures that the input data of each layer has a similar distribution, thereby reducing internal covariate shift. ``LeakyReLU" represents the leaky rectified linear unit activation function, and ``Avg-Pooling" indicates average pooling, which downsamples by computing the average value of a specific region in the previous layer. In the table, ``$a*b$" indicates ``$a$" 1D convolution kernels, each having a size of ``$b$", and ``$p(q)$” represents a fully connected layer with ``$p$" output features and ``$q$" input features.

\subsection{Data Augmentation Methods}

Wideband spectrum sensing based on CNN architectures often requires a large number of training samples to avoid overfitting. However, obtaining a substantial amount of labeled data is challenging in practice, as it demands extensive annotation efforts upfront. Since the spectral information of signals intuitively reflects their occupancy status, we propose two data augmentation methods exploiting frequency-domain information to address the issue of sample scarcity.

\subsubsection{Inter-Subband Shuffle}

The spectrum of communication signals typically consists of multiple overlapping waveforms, but the various sub-bands in the frequency domain usually exhibit a certain degree of independence. This independence characteristic provides the possibility for data augmentation, enabling the effective generation of new samples through the recombination of sub-bands, thereby enhancing the generalization capability of the network.

Specifically, we divide the power spectrum of the entire wideband signal into $N$ subbands:
\begin{equation}
    Y(f) = [{Y_1},{Y_2}, \cdots {Y_i}, \cdots {Y_N}].
\end{equation}
Since the frequency distribution of the signal in different subbands is relatively independent, we can randomize the order of the subbands to generate diversified training samples, while ensuring that the occupancy labels of each subband still correctly reflect its occupancy status. The process is as follows. First, generate a random index array for subband reordering: 
\begin{equation}
    Index = {[4, i - 1, 7, i, 1,  \cdots , i + 1, 3]^N}.
\end{equation}
Then, according to the index array, split and recombine the subbands of the original wideband signal’s power spectrum to form a new wideband signal power spectrum. The reordered subbands are:
\begin{equation}
    \hat Y(f) = [{Y_4}, {Y_{i - 1}}, {Y_7}, {Y_i}, {Y_1},  \cdots , {Y_{i + 1}}, {Y_3}]^N.
\end{equation}

This method reshuffles the order of the subbands. Although the subband data changes, the occupancy labels of each subband remain accurate, ensuring that data diversity is enhanced without affecting the correctness of the labels, thereby improving the performance of the trained model.

 \begin{figure}[t]
    \centering 
    \includegraphics[width=8cm]{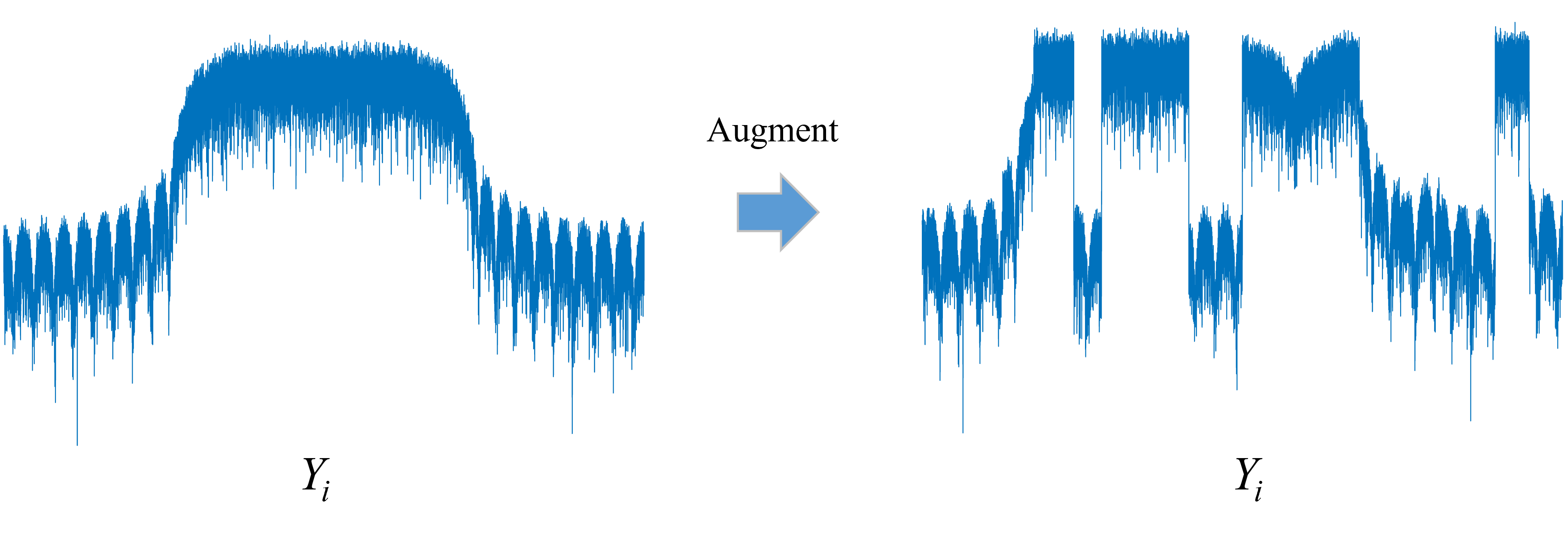}  
    \caption{Intra-subband shuffle enhancement.}
    \label{fig:zq}
\end{figure}
\begin{figure*}[t]
\centering
\subfigure[]{\label{SNR:pd}
\includegraphics[width=0.32\linewidth, height=5.0902cm]{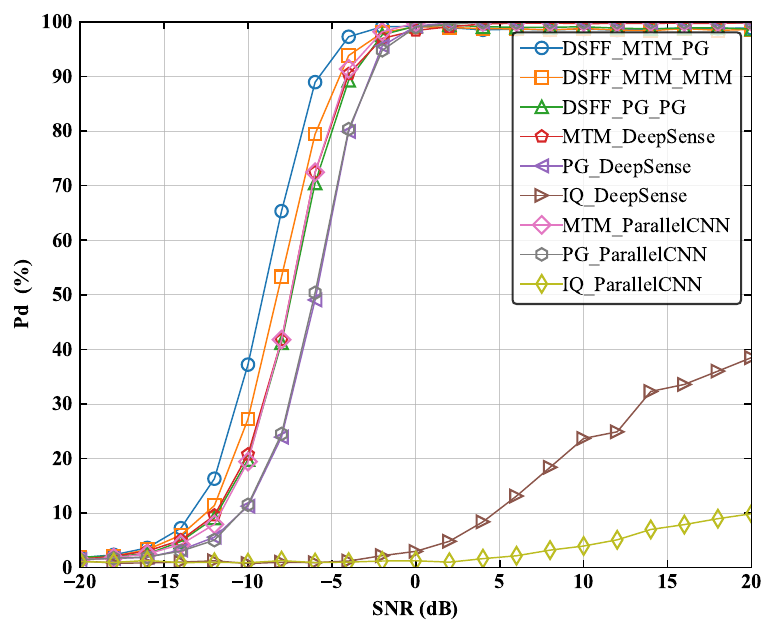}}
\hspace{-1mm}
\subfigure[]{\label{SNR:pf}
\includegraphics[width=0.32\linewidth, height=5.0902cm]{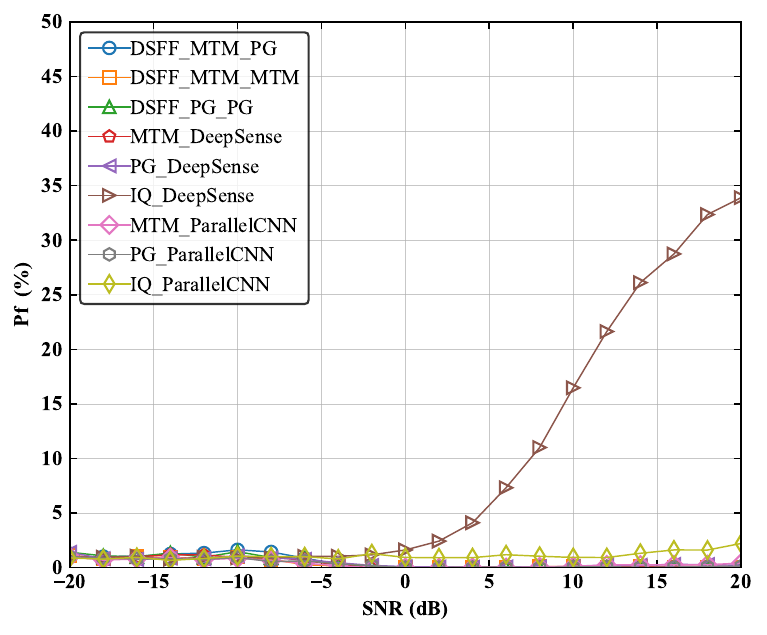}}
\hspace{-1mm}
\subfigure[]{\label{ROC}
\includegraphics[width=0.32\linewidth, height=5.0902cm]{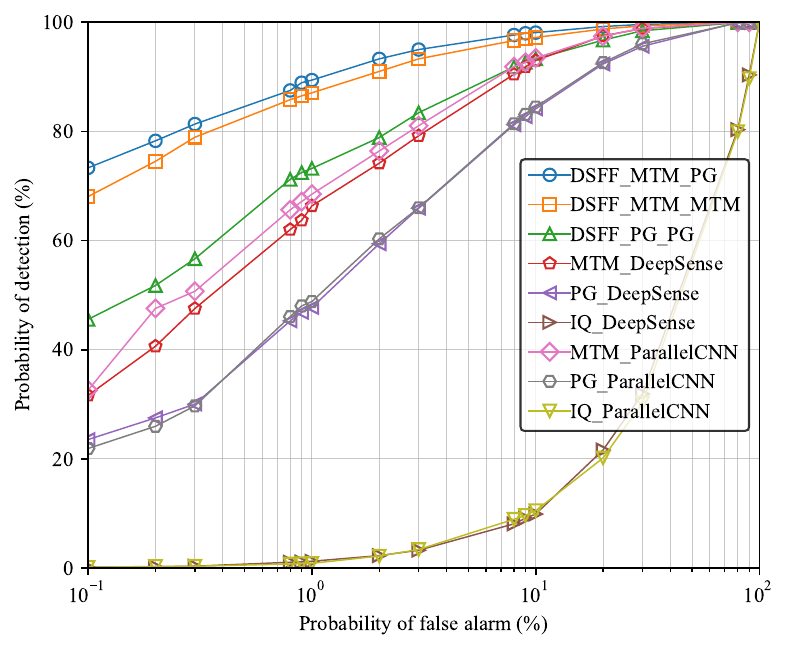}}   
\centering
\caption{Spectrum sensing performance of different algorithms. (a) $P_d$; (b) Actual $P_f$; (c) ROC curve.}
\label{SNR}
\end{figure*}

\subsubsection{Intra-Subband Shuffle}
In practical signal transmission, the primary user's signal typically occupies the central portion of the allocated spectrum, leading to unused frequency bands in certain areas of the spectrum. This can negatively affect deep learning-based spectrum sensing. To address this, we propose the intra-subband shuffle enhancement method, which redistributes the primary user's spectrum occupancy more evenly within a single subband, simulating different spectrum occupancy states. As shown in Fig. \ref{fig:zq}, the intra-subband shuffle enhancement method alters the frequency distribution of the primary user's signal within a single subband by rearranging its spectrum occupancy to form a more uniform distribution. This change breaks the concentration of the original signal, enabling the network to learn more diverse spectral features and improving its ability to adapt to various spectrum occupancy patterns. 

Although data augmentation through Inter-Subband Shuffle and Intra-Subband Shuffle may distort the original spectral structure of each subband and change the order of different primary users' signals, this does not alter the spectral occupancy state of each subband. In other words, the shuffled data still effectively represents different spectral occupancy patterns, and therefore, this impact does not adversely affect the model's performance in spectrum sensing tasks. Moreover, before shuffling, the modulation type of each subband is randomly assigned, ensuring a certain degree of independence between the subbands. The shuffled samples still maintain modulation diversity, allowing the model to adapt to signals with different modulation types. By introducing these two data augmentation methods, we introduce more diverse spectral distribution forms, increasing the diversity of the training data.


\section{Experiment}
\label{sec3}

\subsection{Experimental Setup}

We generate a digital wideband signal dataset that includes multi-user signals, where the entire frequency band is divided into 
$N$ subbands. For each signal generation, the number of users is randomly selected between 1 and 
$N$, and each user’s frequency is assigned randomly to ensure no overlap. The narrowband signals employ a variety of modulation schemes, including BPSK, QPSK, 8PSK, OQPSK, 16PSK, 4PAM, 8PAM, 16QAM, 32QAM, and 64QAM. Pulse shaping is performed using a root-raised cosine filter with a roll-off factor of 0.2, and the symbol duration is randomly chosen from the set {4, 6, 8}.


In the training set, the SNR ranges from 
$-20$ dB to 20 dB in increments of 2 dB, with 100 signal samples generated for each SNR level. To account for the varying SNRs of different users at the receiver, an additional 2,100 samples are generated, where the SNRs for different user signals within the same sample are randomly distributed within the range 
[$-20$, 20] dB. This results in a total of 4,200 samples.

\subsection{Comparison of Different Methods}

We first compare the performance of the method proposed in this paper with those presented in \cite{37,38}, with the experimental results shown in Fig. \ref{SNR}. In this experiment, the false alarm probability is set to 0.01. The notation ``A\_CNN'' is used to represent methods with a single input, where ``A'' indicates the type of input data used by the network, such as IQ data or the power spectrum obtained through MTM or PG, and ``CNN'' denotes the network architecture type. DeepSense and ParallelCNN refer to the network architectures employed in \cite{37} and \cite{38}, respectively. Both the original DeepSense and ParallelCNN methods utilize IQ samples as input. To assess the impact of different input types, we modify the input while keeping the network architecture unchanged. For instance, when the power spectrum obtained through MTM is used as the input, we denote this as ``DeepSense\_MTM''. The notation ``DSFF\_A\_B'' refers to the feature fusion algorithm, where ``DSFF'' represents the feature-fusion-based network architecture proposed in this paper, and ``A'' and ``B'' denote the respective input data types.

The results from the figures show that among single-input algorithms, MTM-based algorithms demonstrate the best performance under both the DeepSense and ParallelCNN architectures. This suggests that compared to PG, MTM more effectively reflects the spectral characteristics of signals, highlighting the advantage of introducing MTM into deep learning. Additionally, power-spectrum-based algorithms outperform IQ-based algorithms across different network architectures. Notably, the proposed DSFF algorithm outperforms all comparative methods, achieving superior detection probability. Compared to PG-based algorithms, the overall performance improves by approximately 3 dB, underscoring the significant advantages of multi-representation input in enhancing spectrum sensing performance.

Specifically, regarding detection probability, as shown in Fig. \ref{SNR:pd}, IQ-based algorithms have a relatively low detection probability across all SNRs due to the limited number of training samples, particularly in low SNR conditions (e.g., below $-5$ dB). In contrast, methods based on power spectrum input show superior detection performance under various SNRs, especially at high SNRs, indicating the advantage of power spectrum input over IQ input. Further comparison of different power spectrum input methods shows that MTM, due to its ability to better extract frequency-domain features, achieves higher detection performance than PG. Compared to single power spectrum input, feature fusion combines information from multiple power spectrum representations, demonstrating superior performance, particularly in low SNR conditions (e.g., below $-5$ dB), underscoring the advantage of the DSFF algorithm in low SNR scenarios.

In terms of false alarm probability, as shown in Fig. \ref{SNR:pf}, despite setting the target false alarm probability to 0.01, it is challenging for IQ-based algorithms to maintain the actual false alarm probability within the target range, especially at high SNR levels. In these cases, the false alarm probability for IQ-based algorithms gradually increases, a phenomenon particularly evident in the DeepSense architecture. In contrast, methods based on power spectrum inputs are more effective in controlling the false alarm probability, keeping it at or below the target level. Overall, the proposed wideband spectrum sensing algorithm not only significantly improves detection probability but also achieves effective control of false alarm probability across the full SNR range, demonstrating superior robustness and reliability.

To compare the performance of different methods at varying false alarm probabilities, we plot the ROC curves of various spectrum sensing methods at an SNR of $-6$ dB, as shown in Fig. \ref{ROC}. The results indicate that the ROC curve of the DSFF-based method is significantly higher than those of other methods, highlighting its superior detection performance across different false alarm probabilities. Specifically, although the detection probability of the proposed DSFF-based algorithm decreases as the false alarm probability is lowered, this decline is less steep compared to other spectrum sensing methods; even at a false alarm probability of 0.001, this method maintains a detection probability above 70\%. This result underscores the advantage of the DSFF-based algorithm in distinguishing between signal and noise. Among single-input algorithms, the detection performance of IQ-based algorithms declines sharply as the false alarm probability decreases, while power-spectrum-based methods exhibit a more gradual decrease. Additionally, the performance of MTM-based algorithms declines more slowly than that of PG-based algorithms, further confirming the effectiveness of MTM in spectrum sensing tasks.

Finally, to verify the effectiveness of power spectrum fusion, we conduct an ablation study comparing the use of dual MTM or dual PG within the feature fusion network. As shown in Fig. \ref{SNR}, the combination of MTM and PG significantly outperforms using either of the power spectrum methods individually, further validating the efficacy of the feature fusion strategy in enhancing spectrum sensing performance.

\subsection{Specific Subband Performance}
\begin{figure}[t]
    \centering 
    \includegraphics[width=1\linewidth]{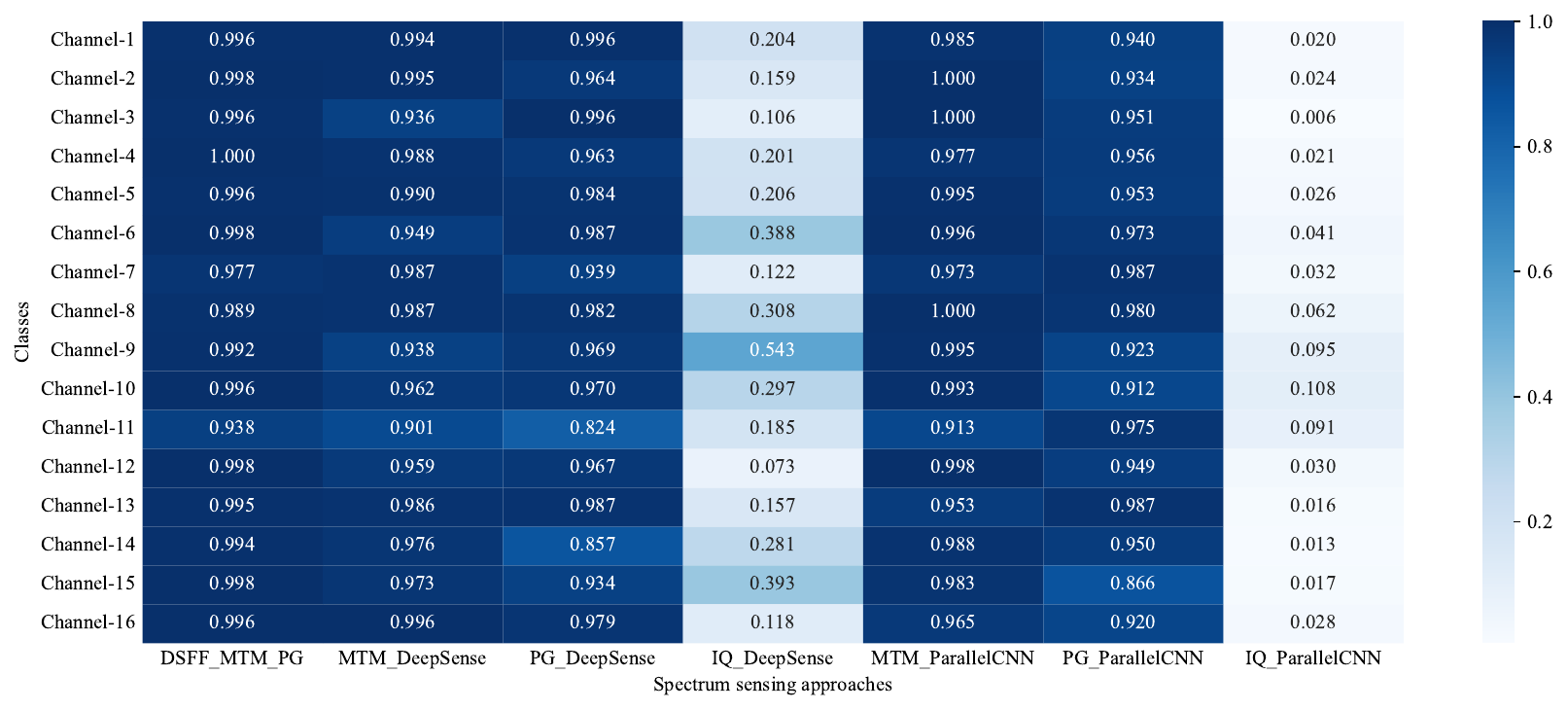}  
    \caption{Spectrum sensing performance of different algorithms in each subchannel.}
    \label{heatmap}
\end{figure}

Fig. \ref{heatmap} illustrates the performance of various spectrum sensing algorithms across subbands at an SNR of $-2$ dB. It can be observed that IQ-based algorithms generally exhibit lower detection probabilities across all subbands compared to power-spectrum-based algorithms, with the ParallelCNN network architecture displaying virtually no detection capability. Among power-spectrum-based algorithms, the PG-based algorithm shows significantly lower detection probabilities in certain subbands, primarily due to signal spectral leakage and filter edge effects. In contrast, the proposed DSFF-based algorithm demonstrates consistently high detection probabilities across all subbands, highlighting its notable advantages in mitigating spectral leakage and edge effects. This result indicates that the feature fusion method can more reliably adapt to the characteristics of different subbands, thereby significantly enhancing overall spectrum sensing performance.

\begin{figure}
\centering
\subfigure[]{
\label{sigNo:pd}
\begin{minipage}[b]{0.22\textwidth}  
\includegraphics[width=1\textwidth]{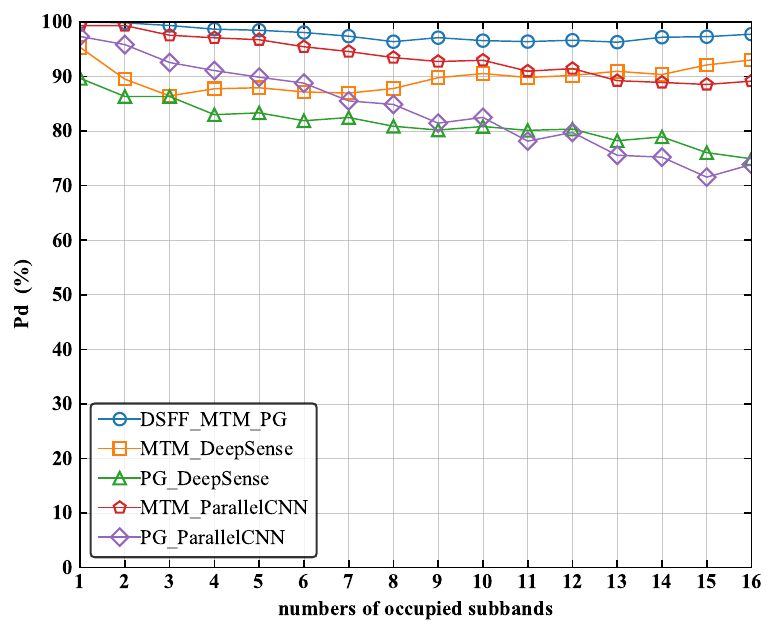} 
\end{minipage}
}
\subfigure[]{
\label{sigNo:pf}
\begin{minipage}[b]{0.22\textwidth}  
\includegraphics[width=1\textwidth]{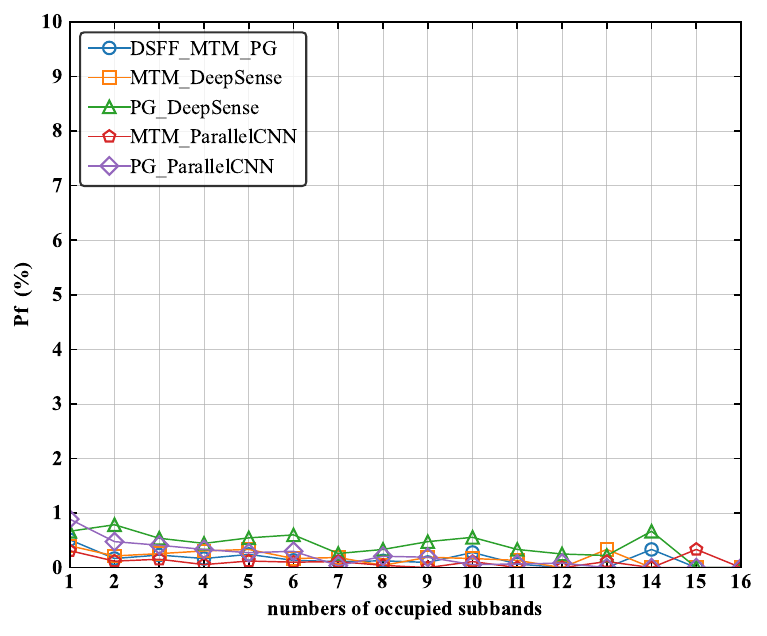} 
\end{minipage}}
\caption{Spectrum sensing performance with different numbers of occupied subbands. (a) $P_d$; (b) Actual $P_f$.}
\label{sigNo}
\end{figure}

\begin{figure*}[t]
\centering
\subfigure[]{
\label{noise_pd}
\includegraphics[width=0.2315\linewidth, height=3.714cm]{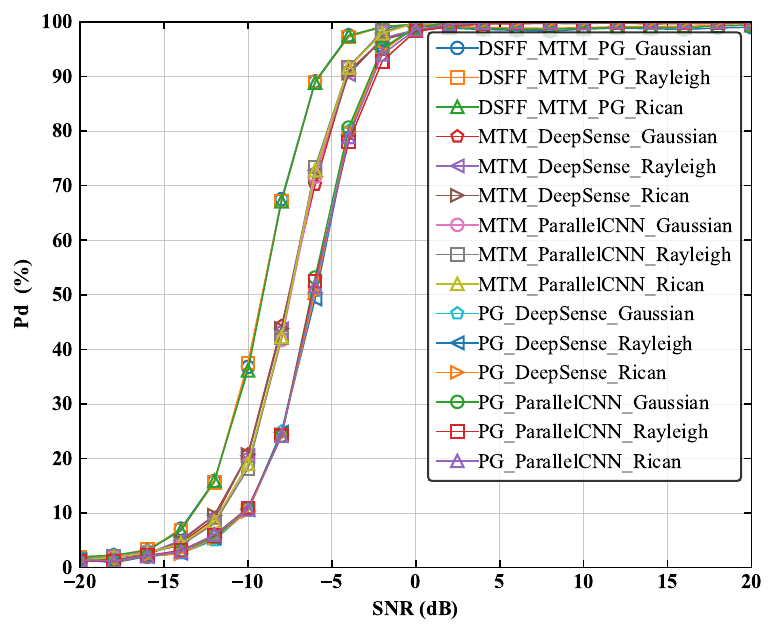}} 
\hspace{-1mm}
\subfigure[]{
\label{noise_pf}
\includegraphics[width=0.2315\linewidth, height=3.714cm]{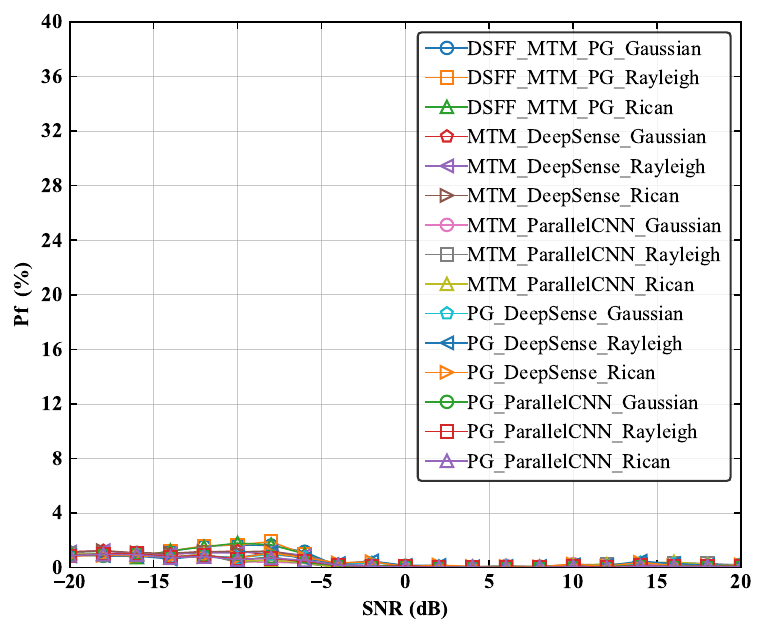}} 
\hspace{-1mm}
\subfigure[]{
\label{noise_IQ_pd}
\includegraphics[width=0.2315\linewidth, height=3.714cm]{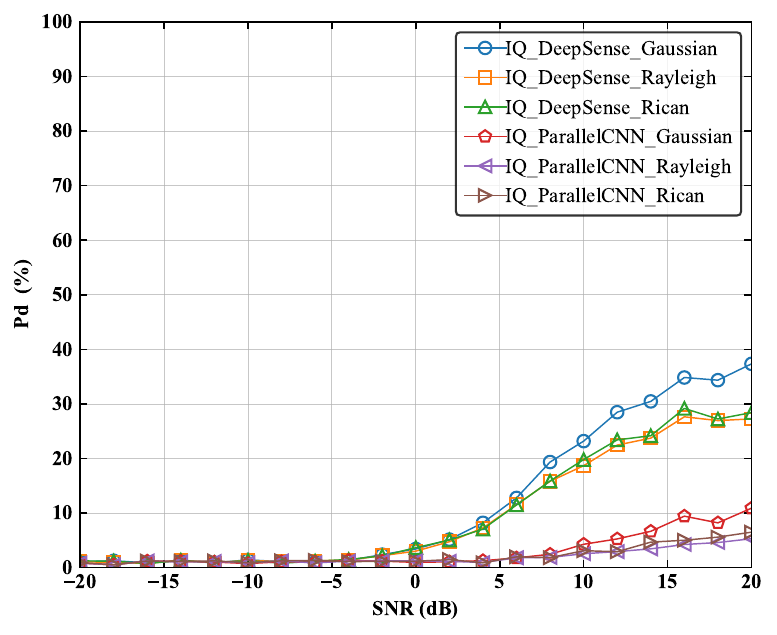}} 
\hspace{-1mm}
\subfigure[]{
\label{noise_IQ_pf}
\includegraphics[width=0.2315\linewidth, height=3.714cm]{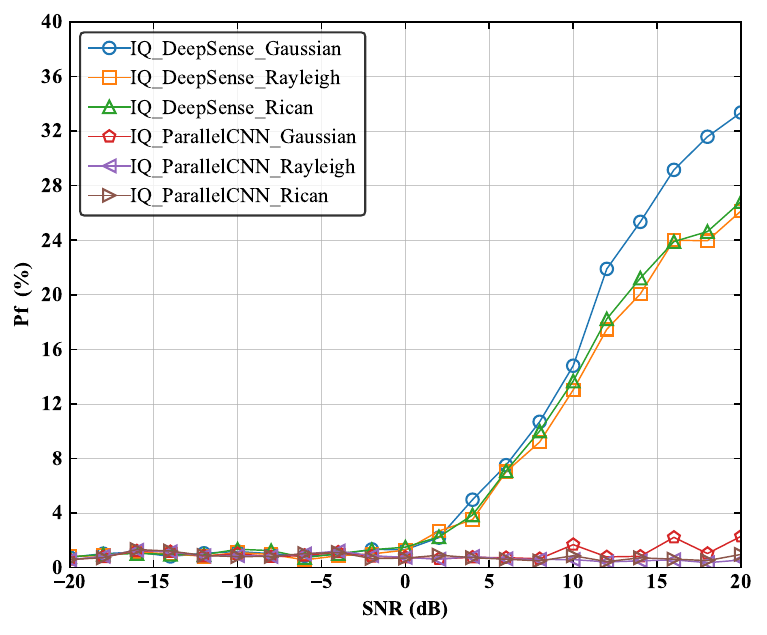}} 
\caption{Spectrum sensing performance in various channel conditions. (a) $P_d$ of PSD\_based algorithms; (b) Actual $P_f$ of PSD\_based algorithms; (c) $P_d$ of IQ\_based algorithms; (d) Actual $P_f$ of IQ\_based algorithms.}
\label{noise}
\end{figure*}

\subsection{Impact of the Number of Occupied Subbands}
To examine the effect of the number of occupied subbands on the performance of various spectrum sensing algorithms, we generate test data at an SNR of $-4$ dB using the same approach as in the generation of experiment data. Fig. \ref{sigNo} shows the impact of different numbers of occupied subbands on detection probability and actual false alarm probability at an SNR of $-4$ dB, with a target false alarm probability of 0.01.


As shown in Fig. \ref{sigNo}, detection performance declines across all algorithms as the number of occupied subbands increases. Nonetheless, the proposed DSFF-based algorithm maintains a high detection probability under varying occupied subband conditions, with actual false alarm probabilities consistently below the target value. This demonstrates its strong robustness and excellent detection performance.

Among single-input algorithms, the power-spectrum-based algorithms exhibit a significant drop in detection probability as the number of occupied subbands rises. In contrast, the MTM-based algorithms achieve more stable performance across different subband occupancy levels. Overall, more effective spectral representation inputs can effectively address high subband occupancy density scenarios.


\begin{figure}[t]
\centering
\subfigure[]{
\label{siglen_deep:pd}
\begin{minipage}[b]{0.22\textwidth}  
\includegraphics[width=1\textwidth]{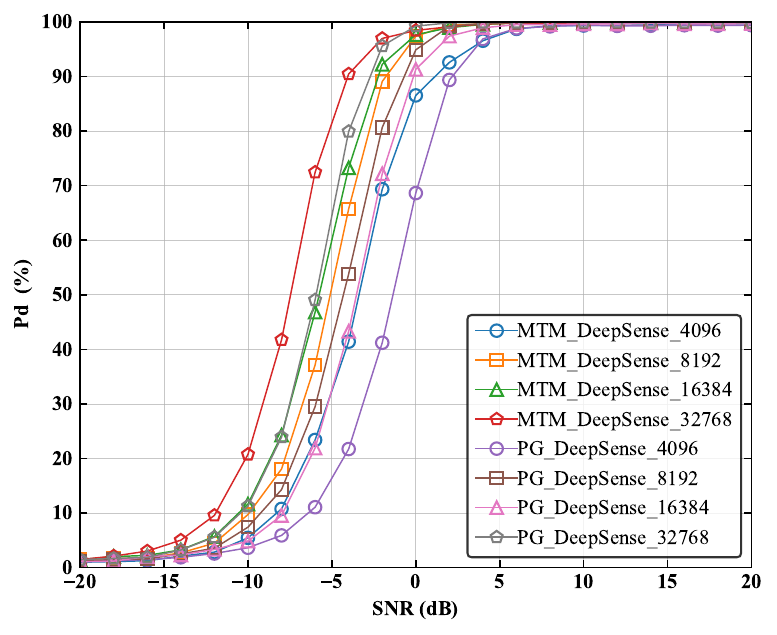} 
\end{minipage}
}
\subfigure[]{
\label{siglen_deep:pf}
\begin{minipage}[b]{0.22\textwidth}  
\includegraphics[width=1\textwidth]{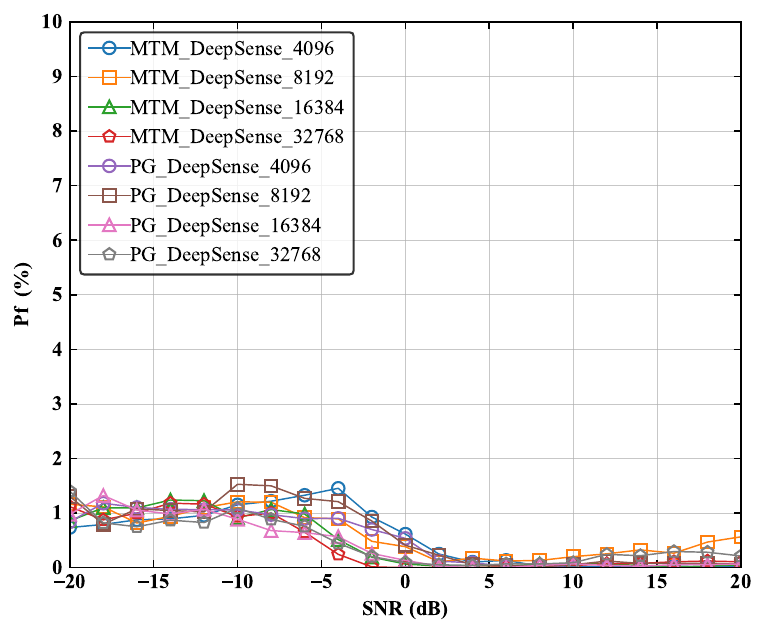} 
\end{minipage}}
\subfigure[]{
\label{siglen_para:pd}
\begin{minipage}[b]{0.22\textwidth}  
\includegraphics[width=1\textwidth]{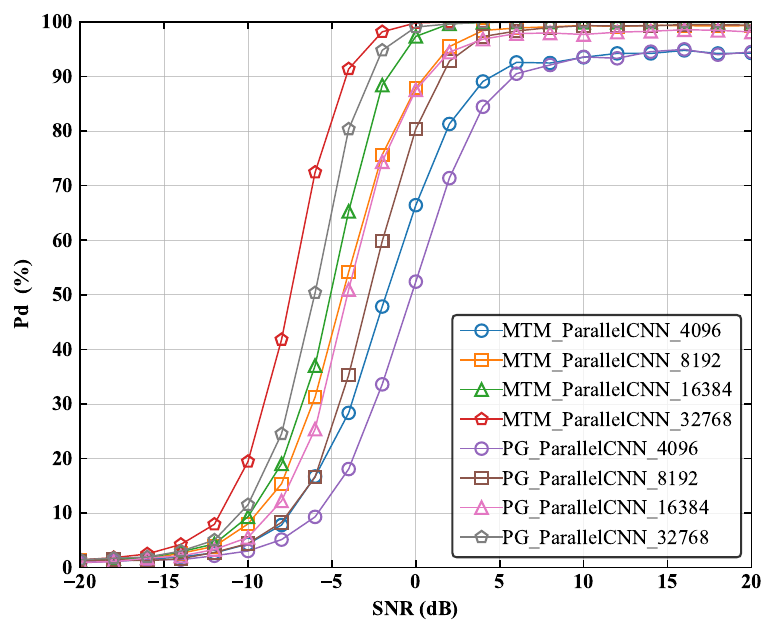} 
\end{minipage}}
\subfigure[]{
\label{siglen_para:pf}
\begin{minipage}[b]{0.22\textwidth}  
\includegraphics[width=1\textwidth]{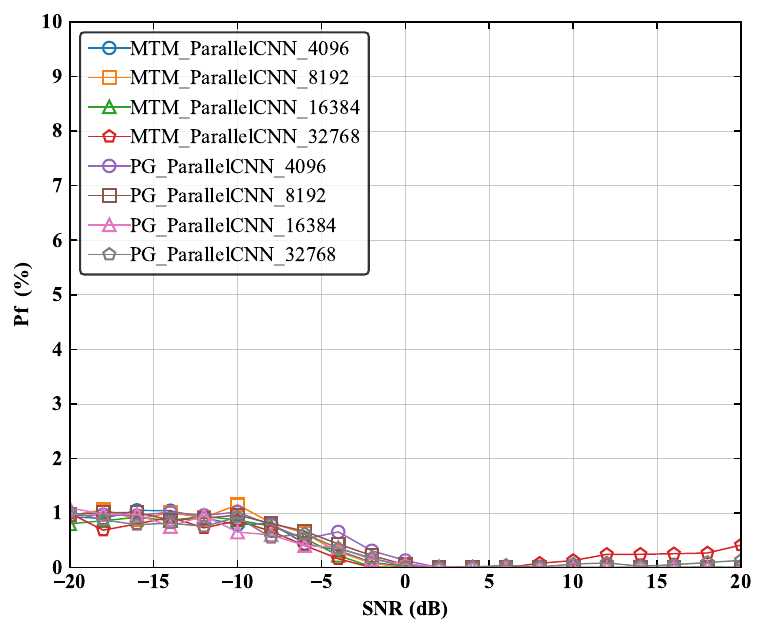} 
\end{minipage}}
\subfigure[]{
\label{siglen_hkdd:pd}
\begin{minipage}[b]{0.22\textwidth}  
\includegraphics[width=1\textwidth]{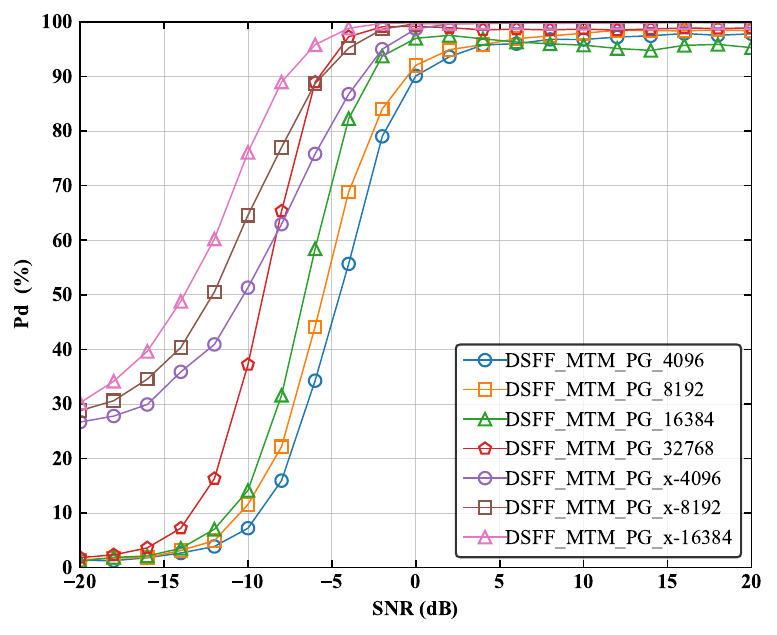} 
\end{minipage}}
\subfigure[]{
\label{siglen_hkdd:pf}
\begin{minipage}[b]{0.22\textwidth}  
\includegraphics[width=1\textwidth]{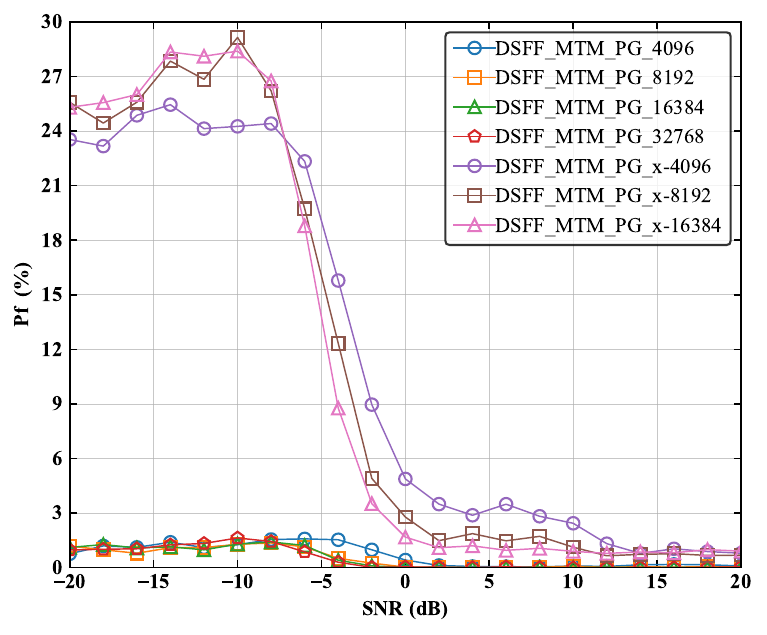} 
\end{minipage}}
\caption{Spectrum sensing performance with different input signal lengths. (a) $P_d$ of DeepSense network model; (b) Actual $P_f$ of DeepSense network model; (c) $P_d$ of the ParallelCNN network model; (d) Actual $P_f$ of the ParallelCNN network model; (e) $P_d$ of the DSFF network model; (f) Actual $P_f$ of the DSFF network model}
\label{siglen}
\end{figure}

\subsection{Performance under Multipath Channel Conditions}
To validate the generalization performance of the algorithms, we further evaluate each algorithm under different channel conditions. We design and generate signals under Rayleigh fading and Rician fading channels, with path delays of [0, 0.5, 1.2] seconds and path gains of [0, $-2$, $-10$] dB and [0, $-5$, $-10$] dB, respectively. Models trained in an AWGN channel environment are used to detect signals in Rayleigh, Rican, and Gaussian channels. Fig. \ref{noise} illustrates the generalization capabilities of various algorithms at a target false alarm probability of 0.01.

From Fig. \ref{noise}, it can be seen that the algorithm based on power spectrum exhibits strong robustness under all channel conditions. This indicates that the power-spectrum-based algorithms have significant adaptability and generalization capability in different channel environments, especially in non-Gaussian channels, and outperform the IQ-based algorithms. Notably, the DSFF-based algorithm proposed in this paper shows almost no variation in detection performance across different channel conditions, consistently maintaining a high detection probability. This result suggests that the DSFF-based algorithm not only has stable adaptability and efficient feature extraction capabilities, but also demonstrates excellent robustness in changing channel environments.


\begin{figure*}[t]
\centering
\subfigure[]{
\label{dj_pd}
\includegraphics[width=0.2315\linewidth, height=3.714cm]{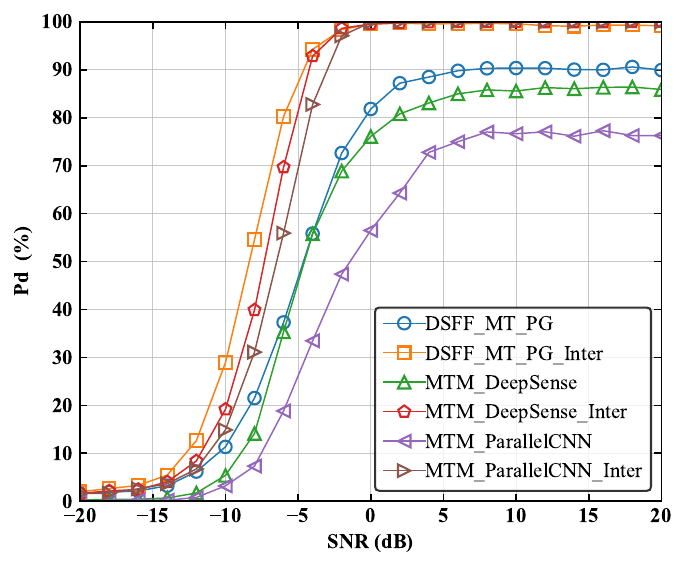}} 
\hspace{-1mm}
\subfigure[]{
\label{dj_pf}
\includegraphics[width=0.2315\linewidth, height=3.714cm]{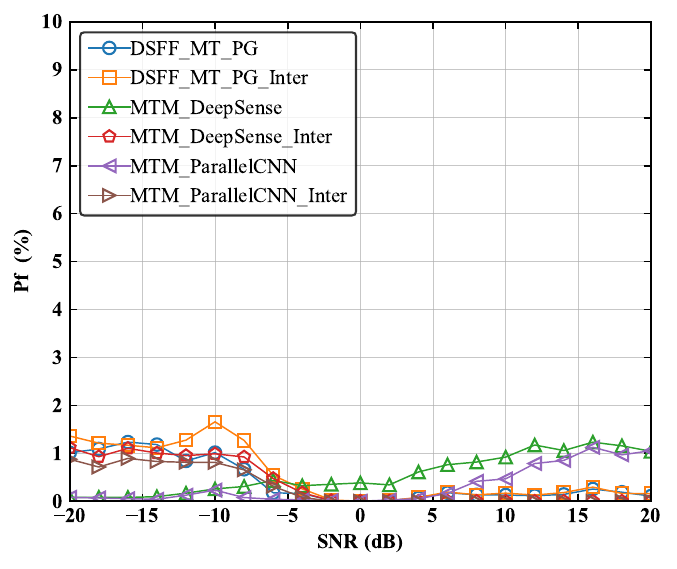}} 
\hspace{-1mm}
\subfigure[]{
\label{dn_pd}
\includegraphics[width=0.2315\linewidth, height=3.714cm]{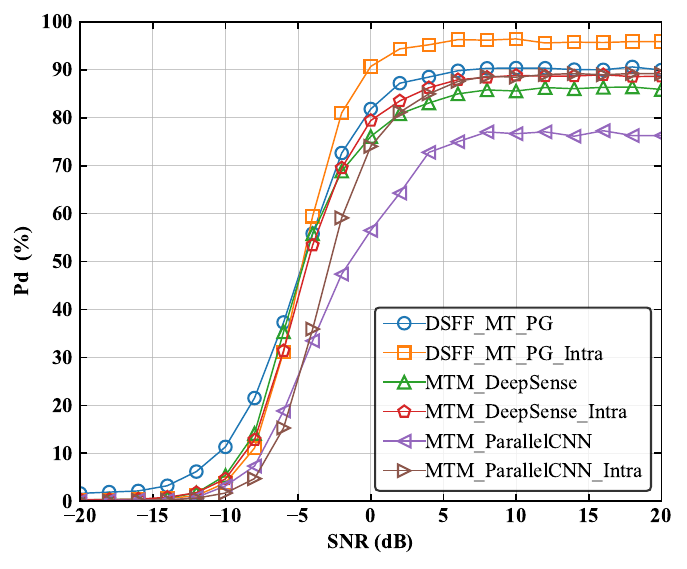}} 
\hspace{-1mm}
\subfigure[]{
\label{dn_pf}
\includegraphics[width=0.2315\linewidth, height=3.714cm]{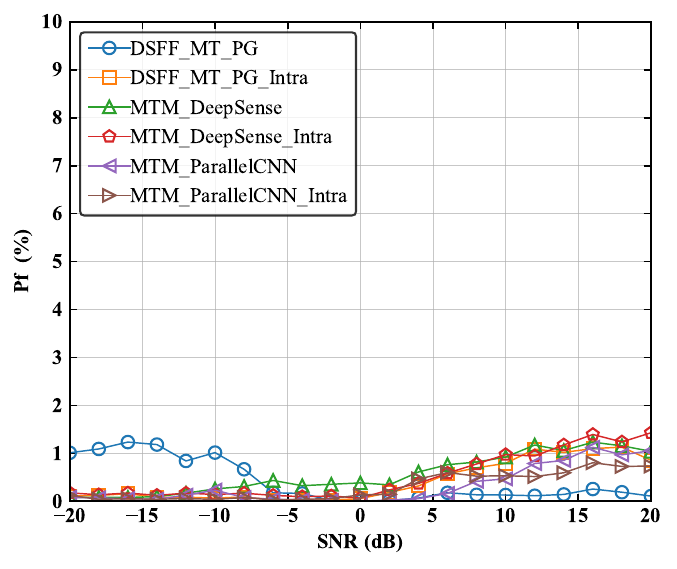}} 
\caption{Spectrum sensing performance under different data augmentation techniques. (a) $P_d$ under inter-subband shuffle; (b) Actual $P_f$ under inter-subband shuffle; (c) $P_d$ under intra-subband shuffle; (d) Actual $P_f$ under intra-subband shuffle.}
\label{enhance}
\end{figure*}
\subsection{Effect of Signal Length}
To evaluate the effect of signal length on the performance of different spectrum sensing algorithms, we generate datasets with different signal lengths and train various networks for performance comparison. The signal lengths used are [32768, 16384, 8192, 4096]. Fig. \ref{siglen} shows the impact of different signal lengths on detection probability and actual false alarm probability under the target false alarm probability of 0.01. Experimental results indicate that as the length of the input signal decreases, the detection probability for all spectrum sensing algorithms decreases. This suggests that longer signal lengths provide more comprehensive spectral information, aiding in the accuracy and stability of detection. As shown in Fig. \ref{siglen}, DSFF-based algorithms and MTM-based algorithms consistently outperform PG-based algorithms across all signal lengths, further validating the effectiveness of MTM for spectrum sensing tasks. In addition, we present the network trained with a signal length of 32768 in Fig. \ref{siglen_hkdd:pd} and \ref{siglen_hkdd:pf}, where different input signal lengths are denoted as ``DSFF\_MTM\_PG\_x-length''. The experimental results demonstrate that, under low SNR conditions, variations in input signal length result in a significant increase in the false alarm probability, along with some meaningless fluctuations in detection probability. In contrast, under high SNR conditions, the detection performance remains robust despite the reduction in input signal length.



\subsection{Performance of Data Augmentation Methods}

We have proposed two data augmentation methods based on frequency-domain shuffling. To evaluate their effectiveness, we extract 420 samples from the training set and apply these methods to augment the data. Each method doubles the original sample size, expanding the dataset to 1,260 samples. The results of the two data augmentation methods, inter-subband shuffle and intra-subband shuffle, are shown in Fig. \ref{enhance}.

\begin{figure}[t]
\centering
\subfigure[]{
\label{both:pd}
\begin{minipage}[b]{0.22\textwidth}  
\includegraphics[width=1\textwidth]{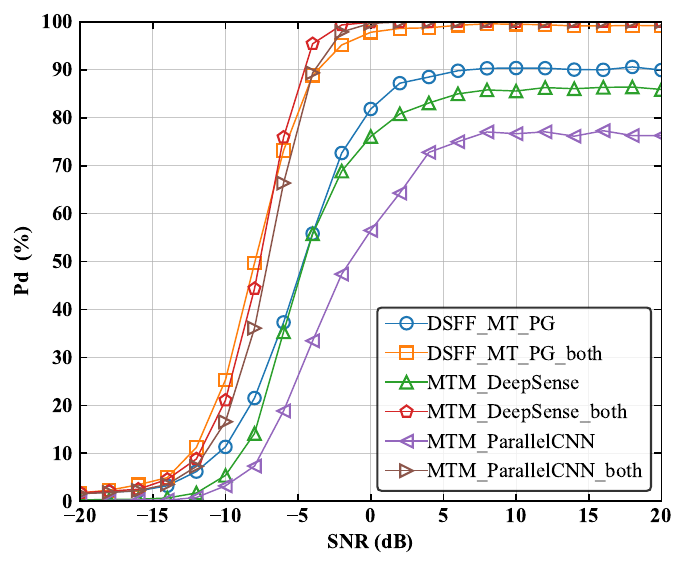} 
\end{minipage}
}
\subfigure[]{
\label{both:pf}
\begin{minipage}[b]{0.22\textwidth}  
\includegraphics[width=1\textwidth]{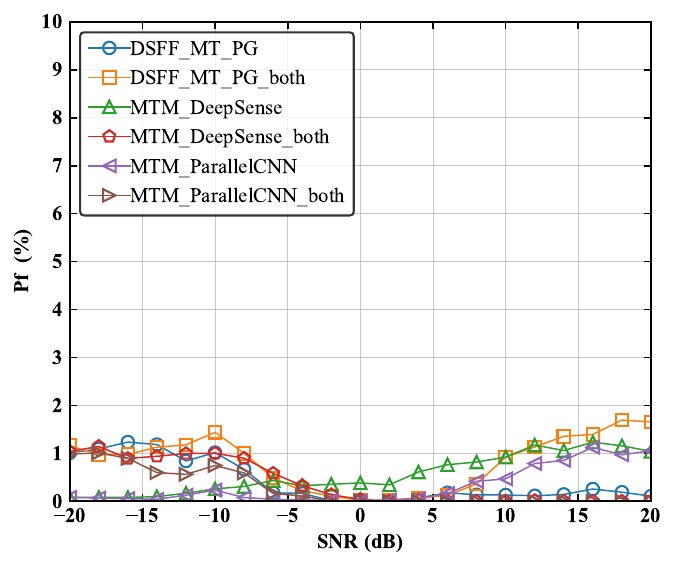} 
\end{minipage}}
\caption{Spectrum sensing performance with the simultaneous use of both data augmentation techniques. (a) $P_d$; (b) Actual $P_f$.}
\label{both}
\end{figure}

Fig. \ref{dj_pd} and \ref{dj_pf} show the model's sensing performance before and after inter-subband shuffle. As can be seen, the detection performance of the model improves significantly after data augmentation. In addition, the experimental results indicate that, when training samples are limited, the proposed DSFF-based algorithm outperforms single-input-based algorithms. Fig. \ref{dn_pd} and \ref{dn_pf} present the model's sensing performance before and after intra-subband shuffle. With fewer samples, the intra-subband shuffle method enhances detection performance, particularly in the DSFF-based and ParallelCNN-based algorithms. Moreover, Fig. \ref{both} shows the results when both augmentation methods are applied simultaneously. Compared to Fig. \ref{enhance}, it is evident that combining both augmentation methods leads to superior sensing performance in the Deepsense-based and ParallelCNN-based algorithms compared to using either method alone. 

It is evident that, when the number of training samples is limited, the detection performance is significantly reduced. However, after applying data augmentation, the accuracy remains close to 100\% at high SNRs, indicating that the proposed data augmentation methods effectively enhance sample utilization efficiency and improve the detection performance of the spectrum sensing algorithm.

\section{Conclusion}
\label{sec4}
This paper presents a novel deep learning-based wideband spectrum sensing framework that enhances spectrum sensing accuracy through dual-representation power spectrum inputs and data augmentation methods designed for few-shot scenarios. Extensive simulations demonstrate that our approach outperforms traditional IQ-based and PG-based algorithms, offering higher detection probability and lower false alarm rates, especially under low SNR conditions. The framework also shows strong generalization and resilience across various channel environments, making it a promising solution for efficient spectrum utilization in cognitive radio networks. Future work will focus on optimizing the method for real-time deployment and exploring additional feature fusion strategies to further improve performance in dynamic environments.

\bibliographystyle{IEEEtran}
\small
\bibliography{ref}

\begin{thebibliography}{10}
\providecommand{\url}[1]{#1}
\csname url@samestyle\endcsname
\providecommand{\newblock}{\relax}
\providecommand{\bibinfo}[2]{#2}
\providecommand{\BIBentrySTDinterwordspacing}{\spaceskip=0pt\relax}
\providecommand{\BIBentryALTinterwordstretchfactor}{4}
\providecommand{\BIBentryALTinterwordspacing}{\spaceskip=\fontdimen2\font plus
\BIBentryALTinterwordstretchfactor\fontdimen3\font minus \fontdimen4\font\relax}
\providecommand{\BIBforeignlanguage}[2]{{%
\expandafter\ifx\csname l@#1\endcsname\relax
\typeout{** WARNING: IEEEtran.bst: No hyphenation pattern has been}%
\typeout{** loaded for the language `#1'. Using the pattern for}%
\typeout{** the default language instead.}%
\else
\language=\csname l@#1\endcsname
\fi
#2}}
\providecommand{\BIBdecl}{\relax}
\BIBdecl

\bibitem{1}
J.~Lunden, V.~Koivunen, and H.~V. Poor, ``Spectrum exploration and exploitation for cognitive radio: Recent advances,'' \emph{IEEE Signal Processing Magazine}, vol.~32, no.~3, pp. 123--140, 2015.

\bibitem{2}
J.~Adu~Ansere, G.~Han, H.~Wang, C.~Choi, and C.~Wu, ``A reliable energy efficient dynamic spectrum sensing for cognitive radio iot networks,'' \emph{IEEE Internet of Things Journal}, vol.~6, no.~4, pp. 6748--6759, 2019.

\bibitem{3}
R.~I. Ansari, H.~Pervaiz, S.~A. Hassan, C.~Chrysostomou, M.~A. Imran, S.~Mumtaz, and R.~Tafazolli, ``A new dimension to spectrum management in iot empowered 5g networks,'' \emph{IEEE Network}, vol.~33, no.~4, pp. 186--193, 2019.

\bibitem{4}
H.~B. Salameh, S.~Otoum, M.~Aloqaily, R.~Derbas, I.~Al~Ridhawi, and Y.~Jararweh, ``Intelligent jamming-aware routing in multi-hop iot-based opportunistic cognitive radio networks,'' \emph{Ad Hoc Networks}, vol.~98, p. 102035, 2020.

\bibitem{5}
Y.~Arjoune and N.~Kaabouch, ``A comprehensive survey on spectrum sensing in cognitive radio networks: Recent advances, new challenges, and future research directions,'' \emph{Sensors}, vol.~19, no.~1, p. 126, 2019.

\bibitem{6}
M.~Sani, J.~Tsado, S.~Thomas, H.~Suleiman, I.~M. Shehu, and M.~G. Shan'una, ``A survey on spectrum sensing techniques for cognitive radio networks,'' in \emph{2021 1st International Conference on Multidisciplinary Engineering and Applied Science (ICMEAS)}, 2021, pp. 1--5.

\bibitem{7}
I.~Mustapha, B.~M. Ali, A.~Sali, M.~F.~A. Rasid, and H.~Mohamad, ``An energy efficient reinforcement learning based cooperative channel sensing for cognitive radio sensor networks,'' \emph{Pervasive and Mobile Computing}, vol.~35, pp. 165--184, 2017.

\bibitem{8}
I.~Raghu, S.~S. Chowdary, and E.~Elias, ``Efficient spectrum sensing for cognitive radio using cosine modulated filter banks,'' in \emph{2016 IEEE Region 10 Conference (TENCON)}, 2016, pp. 2086--2089.

\bibitem{9}
K.~Sharma and A.~Sharma, ``Design of cosine modulated filter banks exploiting spline function for spectrum sensing in cognitive radio applications,'' in \emph{2016 IEEE 1st International Conference on Power Electronics, Intelligent Control and Energy Systems (ICPEICES)}, 2016, pp. 1--5.

\bibitem{10}
Z.~Quan, S.~Cui, A.~H. Sayed, and H.~V. Poor, ``Wideband spectrum sensing in cognitive radio networks,'' in \emph{2008 IEEE International Conference on Communications}, 2008, pp. 901--906.

\bibitem{11}
------, ``Optimal multiband joint detection for spectrum sensing in cognitive radio networks,'' \emph{IEEE Transactions on Signal Processing}, vol.~57, no.~3, pp. 1128--1140, 2009.

\bibitem{12}
Z.~Tian, Y.~Tafesse, and B.~M. Sadler, ``Cyclic feature detection with sub-nyquist sampling for wideband spectrum sensing,'' \emph{IEEE Journal of Selected Topics in Signal Processing}, vol.~6, no.~1, pp. 58--69, 2012.

\bibitem{13}
Z.~Tian and G.~B. Giannakis, ``A wavelet approach to wideband spectrum sensing for cognitive radios,'' in \emph{2006 1st International Conference on Cognitive Radio Oriented Wireless Networks and Communications}, 2006, pp. 1--5.

\bibitem{14}
E.~J. Candes and T.~Tao, ``Near-optimal signal recovery from random projections: Universal encoding strategies?'' \emph{IEEE Transactions on Information Theory}, vol.~52, no.~12, pp. 5406--5425, 2006.

\bibitem{15}
Z.~Tian and G.~B. Giannakis, ``Compressed sensing for wideband cognitive radios,'' in \emph{2007 IEEE International Conference on Acoustics, Speech and Signal Processing - ICASSP '07}, vol.~4, 2007, pp. IV--1357--IV--1360.

\bibitem{16}
R.~G. Baraniuk, ``Compressive sensing [lecture notes],'' \emph{IEEE Signal Processing Magazine}, vol.~24, no.~4, pp. 118--121, 2007.

\bibitem{17}
S.~Kirolos, J.~Laska, M.~Wakin, M.~Duarte, D.~Baron, T.~Ragheb, Y.~Massoud, and R.~Baraniuk, ``Analog-to-information conversion via random demodulation,'' in \emph{2006 IEEE Dallas/CAS Workshop on Design, Applications, Integration and Software}, 2006, pp. 71--74.

\bibitem{18}
M.~Mishali and Y.~C. Eldar, ``From theory to practice: Sub-nyquist sampling of sparse wideband analog signals,'' \emph{IEEE Journal of Selected Topics in Signal Processing}, vol.~4, no.~2, pp. 375--391, 2010.

\bibitem{19}
------, ``Blind multiband signal reconstruction: Compressed sensing for analog signals,'' \emph{IEEE Transactions on Signal Processing}, vol.~57, no.~3, pp. 993--1009, 2009.

\bibitem{20}
Z.~Qin, Y.~Gao, M.~D. Plumbley, and C.~G. Parini, ``Wideband spectrum sensing on real-time signals at sub-nyquist sampling rates in single and cooperative multiple nodes,'' \emph{IEEE Transactions on Signal Processing}, vol.~64, no.~12, pp. 3106--3117, 2016.

\bibitem{21}
B.~Khalfi, B.~Hamdaoui, M.~Guizani, and N.~Zorba, ``Efficient spectrum availability information recovery for wideband dsa networks: A weighted compressive sampling approach,'' \emph{IEEE Transactions on Wireless Communications}, vol.~17, no.~4, pp. 2162--2172, 2018.

\bibitem{22}
L.~Yang, J.~Fang, H.~Duan, and H.~Li, ``Fast compressed power spectrum estimation: Toward a practical solution for wideband spectrum sensing,'' \emph{IEEE Transactions on Wireless Communications}, vol.~19, no.~1, pp. 520--532, 2020.

\bibitem{23}
F.~Wang, J.~Fang, H.~Duan, and H.~Li, ``Phased-array-based sub-nyquist sampling for joint wideband spectrum sensing and direction-of-arrival estimation,'' \emph{IEEE Transactions on Signal Processing}, vol.~66, no.~23, pp. 6110--6123, 2018.

\bibitem{24}
E.~Lagunas and M.~Nájar, ``Spectral feature detection with sub-nyquist sampling for wideband spectrum sensing,'' \emph{IEEE Transactions on Wireless Communications}, vol.~14, no.~7, pp. 3978--3990, 2015.

\bibitem{25}
D.~Romero, D.~D. Ariananda, Z.~Tian, and G.~Leus, ``Compressive covariance sensing: Structure-based compressive sensing beyond sparsity,'' \emph{IEEE Signal Processing Magazine}, vol.~33, no.~1, pp. 78--93, 2016.

\bibitem{26}
S.~Zheng, S.~Chen, and X.~Yang, ``Deepreceiver: A deep learning-based intelligent receiver for wireless communications in the physical layer,'' \emph{IEEE Transactions on Cognitive Communications and Networking}, vol.~7, no.~1, pp. 5--20, 2021.

\bibitem{27}
S.~Zheng, X.~Zhou, L.~Zhang, P.~Qi, K.~Qiu, J.~Zhu, and X.~Yang, ``Toward next-generation signal intelligence: A hybrid knowledge and data-driven deep learning framework for radio signal classification,'' \emph{IEEE Transactions on Cognitive Communications and Networking}, vol.~9, no.~3, pp. 564--579, 2023.

\bibitem{28}
S.~Zheng, Z.~Yang, W.~Shen, L.~Zhang, J.~Zhu, Z.~Zhao, and X.~Yang, ``Deep learning-based doa estimation,'' \emph{IEEE Transactions on Cognitive Communications and Networking}, vol.~10, no.~3, pp. 819--835, 2024.

\bibitem{29}
H.~Xing, H.~Qin, S.~Luo, P.~Dai, L.~Xu, and X.~Cheng, ``Spectrum sensing in cognitive radio: A deep learning based model,'' \emph{Transactions on Emerging Telecommunications Technologies}, vol.~33, no.~1, p. e4388, 2022.

\bibitem{30}
S.~Zheng, S.~Chen, P.~Qi, H.~Zhou, and X.~Yang, ``Spectrum sensing based on deep learning classification for cognitive radios,'' \emph{China Communications}, vol.~17, no.~2, pp. 138--148, 2020.

\bibitem{31}
G.~Pan, J.~Li, and F.~Lin, ``A cognitive radio spectrum sensing method for an ofdm signal based on deep learning and cycle spectrum,'' \emph{International Journal of Digital Multimedia Broadcasting}, vol. 2020, no.~1, p. 5069021, 2020.

\bibitem{32}
A.~Vagollari, V.~Schram, W.~Wicke, M.~Hirschbeck, and W.~Gerstacker, ``Joint detection and classification of rf signals using deep learning,'' in \emph{2021 IEEE 93rd Vehicular Technology Conference (VTC2021-Spring)}, 2021, pp. 1--7.

\bibitem{33}
N.~Ambika, K.~Muthumeenakshi, and S.~Radha, ``Classification of primary user occupancy using deep learning technique in cognitive radio,'' in \emph{Advances in Automation, Signal Processing, Instrumentation, and Control: Select Proceedings of i-CASIC 2020}.\hskip 1em plus 0.5em minus 0.4em\relax Springer, 2021, pp. 1795--1804.

\bibitem{34}
X.~Zha, H.~Peng, X.~Qin, G.~Li, and S.~Yang, ``A deep learning framework for signal detection and modulation classification,'' \emph{Sensors}, vol.~19, no.~18, p. 4042, 2019.

\bibitem{35}
Y.~Zhang, B.~Shen, J.~Wang, and T.~Yan, ``Cnn based wideband spectrum occupancy status identification for cognitive radios,'' in \emph{2020 International Conference on Wireless Communications and Signal Processing (WCSP)}, 2020, pp. 569--574.

\bibitem{36}
W.~Zhang, Y.~Wang, X.~Chen, Z.~Cai, and Z.~Tian, ``Spectrum transformer: An attention-based wideband spectrum detector,'' \emph{IEEE Transactions on Wireless Communications}, vol.~23, no.~9, pp. 12\,343--12\,353, 2024.

\bibitem{37}
D.~Uvaydov, S.~D’Oro, F.~Restuccia, and T.~Melodia, ``Deepsense: Fast wideband spectrum sensing through real-time in-the-loop deep learning,'' in \emph{IEEE INFOCOM 2021 - IEEE Conference on Computer Communications}, 2021, pp. 1--10.

\bibitem{38}
R.~Mei and Z.~Wang, ``Deep learning-based wideband spectrum sensing: A low computational complexity approach,'' \emph{IEEE Communications Letters}, vol.~27, no.~10, pp. 2633--2637, 2023.

\end{thebibliography}

 \begin{IEEEbiography}[{\includegraphics[width=1in,height=1.25in,clip,keepaspectratio]{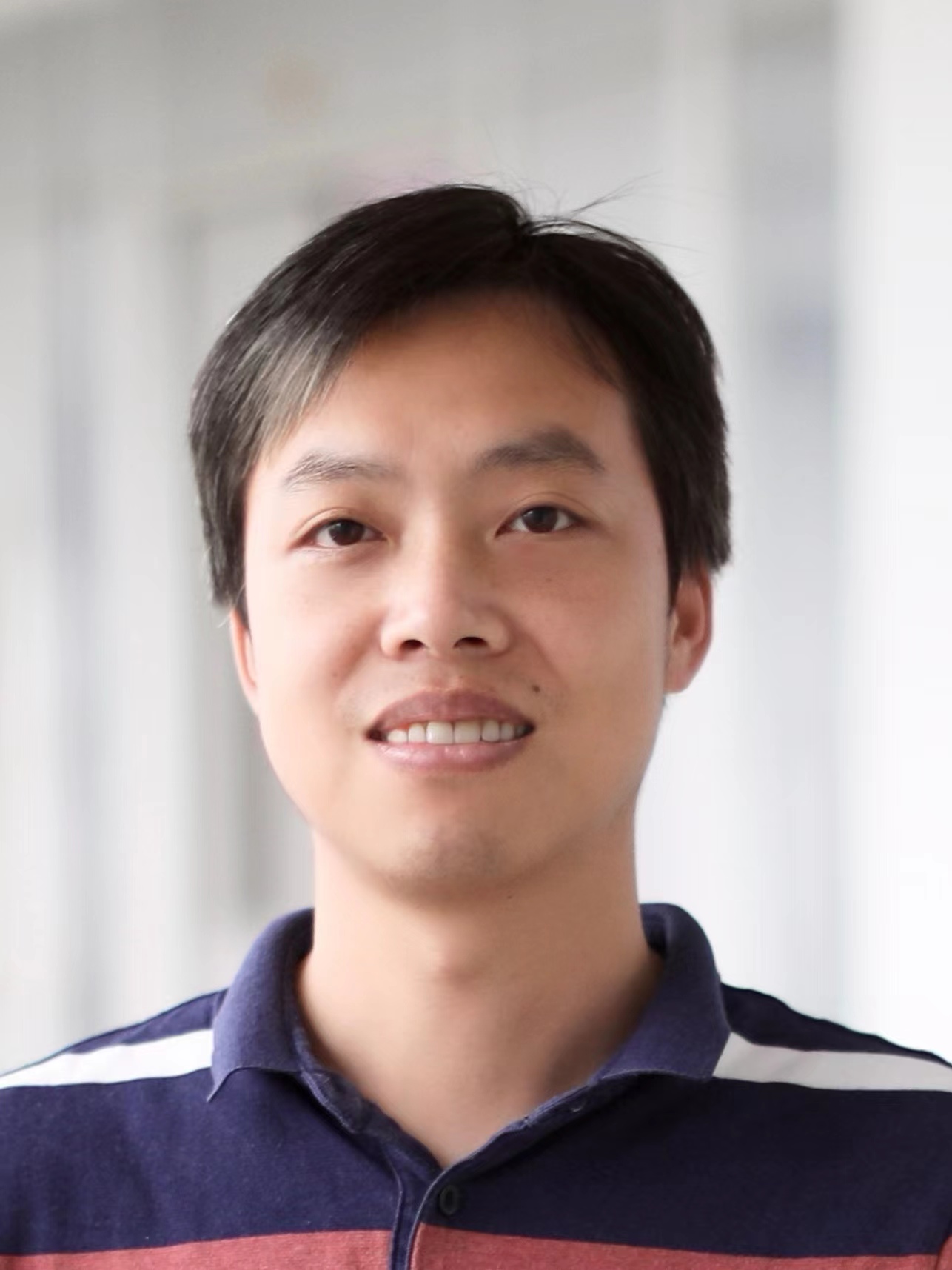}}]{Shilian Zheng}
 received the B.S. degree in telecommunication engineering and the M.S. degree in signal and information processing from Hangzhou Dianzi University, Hangzhou, China, in 2005 and 2008, respectively, and the Ph.D. degree in communication and information system from Xidian University, Xi'an, China, in 2014.

 He is currently a Researcher with National Key Laboratory of Electromagentic Space Security, Jiaxing, China, and a Doctoral Supervisor at Hangzhou Dianzi University, Hangzhou, China. His research interests include cognitive radio, deep learning-based radio signal processing, and electromagnetic space security.
 \end{IEEEbiography}

 \begin{IEEEbiography}[{\includegraphics[width=1in,height=1.25in,clip,keepaspectratio]{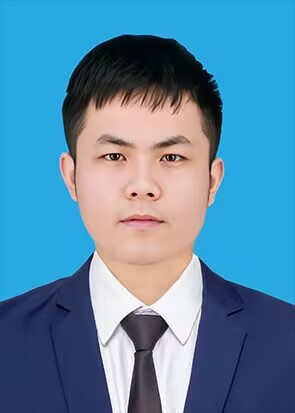}}]{Zhihao Ye}
received the B.S. degree in communications engineering from Hangzhou Dianzi University, Hangzhou, China, in 2023. He is currently pursuing the M.S. degree in communications engineering with the School of Communications Engineering, Hangzhou Dianzi University, Hangzhou. His current research interests include deep learning and wideband spectrum sensing.
 \end{IEEEbiography}



\begin{IEEEbiography}[{\includegraphics[width=1in,height=1.25in,clip,keepaspectratio]{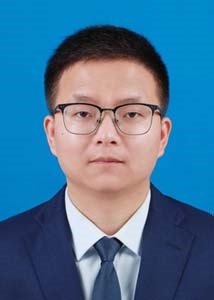}}]{Luxin Zhang}
received the M.S. degree in control science and engineering from Zhejiang University of Technology, Hangzhou, China, in 2021. 

He is currently an Assistant Engineer with National Key Laboratory of Electromagentic Space Security, Jiaxing, China. His research interests include cognitive radio, radio signal processing and learning-based radio signal recognition.
\end{IEEEbiography}

\begin{IEEEbiography}[{\includegraphics[width=1in,height=1.25in,clip,keepaspectratio]{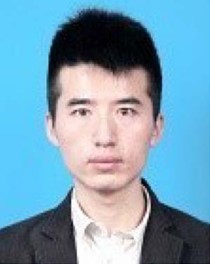}}]{Keqiang Yue}
received the B.E. degree from Anyang Normal University, Anyang, China, in 2007, the B.E. degree from Hangzhou Dianzi University, Hangzhou, China, in 2010, and the Ph.D. degree from Zhejiang University, Hangzhou, China, in 2014. He is currently an Associate Professor with the Hangzhou Dianzi University. His research interests include deep learning based vital signs process and communication signal processing.
\end{IEEEbiography}


\begin{IEEEbiography}[{\includegraphics[width=1in,height=1.25in,clip,keepaspectratio]{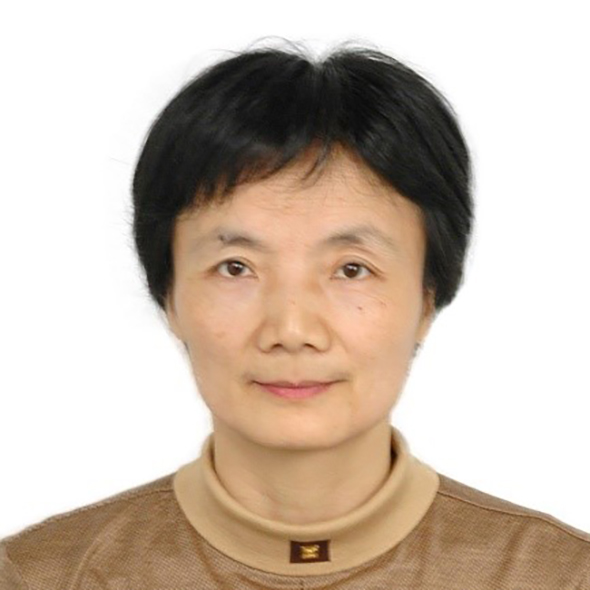}}]{Zhijin Zhao}
received the M.S. and Ph.D. degrees from from Xidian University, Xi’an, China, in 1984 and 2009, respectively. In 1993 and 2003, as a Visiting Scholar, she studied adaptive signal processing and blind signal processing in Darmstadt University of Technology and University of Erlangen-Nuremberg respectively. She is currently a Professor with the School of Communication Engineering, Hangzhou Dianzi University, Hangzhou, China. Her research interests include communication signal processing, cognitive radio technology, intelligent signal processing, and other aspects of research. She once served as President of the School of Communication Engineering, Hangzhou Dianzi University, as well as a Senior Member of China Electronics Society, a member of National Signal Processing Society.
\end{IEEEbiography}



\end{document}